\documentclass[
showkeys,	%	Control display of Keywords: line.
preprint	%	Single-column with increased spacing.
raggedbottom,	%Control [twocolumn] balancing.
flushbottom,
amsmath,	%	Load amsmath package.- AMS-LATEX features.
amssymb,	%	Load amssymb package.-  AMS symbols.
amsfonts,	%	Load amsfonts package.-  AMS font support.
aps,		%	American Physical Society styling.
pre,%
groupedaddress,% Group authors with same affiliations.
superscriptaddress,% Associate authors with 			affiliations via superscript numbers. Appropriate for collaborations or if several authors share some, but not all, affiliations.
a4paper,%
nofootinbib,%	Control location of footnotes.
floatfix,	%
preprintnumbers, % Control display of preprint numbers given by \preprint command.
]{revtex4-2}

\usepackage{amsthm,mathtools}

\usepackage{subfigure}

\usepackage{multirow}
\usepackage{physics}
\usepackage{dcolumn}% Align table columns on decimal point
\usepackage{bm}% bold math
\usepackage{graphics,graphicx} 
\usepackage{wasysym}%,stix}%geometric shapes

\usepackage{xcolor, color}
\definecolor{BlueGreen}{RGB}{0, 154, 166}      
\definecolor{VioletRed}{RGB}{215, 31, 133}     
\definecolor{Purple}{RGB}{182, 52, 187} 	
\definecolor{Turquoise}{RGB}{0, 255, 239} 	   
\definecolor{amber}{rgb}{1.0, 0.75, 0.0}
\definecolor{uscgold}{rgb}{1.0, 0.8, 0.0}
\definecolor{uclagold}{rgb}{1.0, 0.7, 0.0}
\definecolor{gold(metallic)}{rgb}{0.83, 0.69, 0.22}
\definecolor{gold(web)(golden)}{rgb}{1.0, 0.84, 0.0}
\definecolor{goldenpoppy}{rgb}{0.99, 0.76, 0.0}
\definecolor{goldenyellow}{rgb}{1.0, 0.87, 0.0}
\definecolor{goldenrod}{rgb}{0.85, 0.65, 0.13}	   

\usepackage[%backref=page
]{hyperref}%reference back to page

\usepackage{natbib}
\hypersetup{
colorlinks=true,
linkcolor={red},
filecolor=blue,      
urlcolor={BlueGreen},
citecolor=cyan,
anchorcolor=orange,
linktocpage=true,
breaklinks,
breaklinks=true,
final
}

\usepackage{booktabs,graphicx}
\def\sym#1{\ifmmode^{#1}\else\(^{#1}\)\fi}

\usepackage{siunitx}

\usepackage{float}
\usepackage{tabularray}

\usepackage{soul, pifont}

\usepackage[capitalise]{cleveref}
\crefname{figure}{FIG.}{FIGS.}
\Crefname{figure}{FIG.}{FIGS.} % Handles \Cref at start of sentence
\makeatletter
\AtBeginDocument
{
\def\ltx@label#1{\cref@label{#1}}%add braces
\def\label@in@display@noarg#1{\cref@old@label@in@display{#1}}%remove braces
\def\label@in@mmeasure@noarg#1{%
	\begingroup%
	\measuring@false%
	\cref@old@label@in@display{#1}%remove braces for multline, see https://tex.stackexchange.com/q/737204/2388
	\endgroup}%  
	} %
	\makeatother
	
	\usepackage{enumitem}
	
	\usepackage{orcidlink}
	\graphicspath{{Figures/}} % Directory in which figures are stored
\begin{document}

%\preprint{TANU/DRAFT-3}
%	\preprint{XYZ}		
\normalsize
%\title{Chaos/Butterfly effect in Gaussian Perturbed Harmonic Potential well\\
	%(OR)The Role of Random Perturbations in Triggering Chaos in a Harmonic System\\
	%(OR)Small Perturbations, Large Effects: Chaos in a Harmonic potential well with Gaussian Noise\\
	%(OR)Gaussian Noise-Induced Transition to Chaos in a Harmonic potential well\\
	%(OR)Exploring Sensitivity and Unpredictability in a Noisy Harmonic System\\
	%(OR)Noise-Driven Unpredictability in a Harmonic Potential well\\
	%(OR)Stochastic Sensitivity and the Emergence of Chaos in a Harmonic potential well\\}
\title{%Emergence of chaos in cymatics-inspired topographies with localised Gaussian bump/dimple distributions\\or\\
	Chaos in cymatics-inspired Gaussian landscapes}

%\thanks{A bump (dimple) denotes a positive (negative) Gaussian perturbation that locally elevates (lowers) the harmonic potential while preserving smoothness and differentiability.}%A footnote to the article title

\author{Tanmayee Patra\,\orcidlink{0009-0007-5555-0068}}
\email{tanmayeepatra1995@gmail.com}
\affiliation{%
	%\normalfont
	Department of Physics and Astronomy, National Institute of Technology Rourkela, Odisha-769008, India}%

\author{Pranaya Pratik Das\,\orcidlink{0000-0002-6025-7719}}
\email{pranaya.phy@outlook.com}
%\email{pranayapratik_das@nitrkl.ac.in}
\affiliation{%
	%\normalfont
	Department of Physics and Astronomy, National Institute of Technology Rourkela, Odisha-769008, India}%

\author{Biplab Ganguli\,\orcidlink{0000-0003-2583-5752}}%
\email{biplabg@nitrkl.ac.in}
\affiliation{%
	%\normalfont
	Department of Physics and Astronomy, National Institute of Technology Rourkela, Odisha-769008, India}%

\begin{abstract}
	This paper presents a focused investigation of a conservative chaotic system, specifically within the context of a two-dimensional harmonic potential well. We analyse the emergence of chaos from a straightforward, non-chaotic harmonic potential well when subjected to perturbations introduced by two Gaussian-like terms in the system's Hamiltonian. The Gaussian-perturbed system serves as a foundation for further inquiries rooted in the cymatics mechanism. In this study, we examine the effects of deformations arising from Gaussian perturbations on the development of chaotic dynamics. These deformations are produced through various configurations of Gaussian bumps in different geometric shapes, along with the modulation of the amplitude of the perturbed term shifting from positive to negative values.
	
\end{abstract}

\maketitle
\pagestyle{plain}
%\tableofcontents
%%%%%%%%%%%%%%%%%%%%%%%%%%%%%%%%%%%%%%%%%%%%%%%%%%%
\section{\label{secI}Introduction}
%\section{\label{sec1}i\lowercase{ntroduction}}
%%%%%%%%%%%%%%%%%%%%%%%%%%%%%%%%%%%%%%%%%%%%%%%

%The most basic technique of studying any nonlinear system is to study phase space portrait\cite{strogatz2001nonlinear}.
%\cite{ Azc_rraga_1997, RevModPhys.53.655, jensen1987classical, hayes1991introduction,Hsiao2017IntroductionTC}.

%%%%%%%%%%%%%%%%%%%%%%%%%%
Since the foundational works of Henri Poincaré (three body problem) \cite{poincare2017three}, the relationship between dynamical evolution and geometric structure lies at the heart of classical mechanics. A persistent question arising from this tradition is how strictly deterministic laws can generate motion that is aperiodic and highly sensitive to initial conditions. This apparent paradox finds its most precise formulation in the theory of Hamiltonian systems. In Hamiltonian systems, geometry serves as more than just a descriptive tool; it acts as a structural constraint. The symplectic form of Hamiltonian systems preserves the volume of phase space, while invariant manifolds organize motion in the system. Additionally, the stability of the system is reflected in the topology of tori embedded within energy surfaces. From this viewpoint, chaos does not arise from randomness but rather due to a geometric instability\textemdash the gradual deformation and eventual destruction of invariant structures under influence of perturbation.

The study of Hamiltonian chaos has historically progressed along two parallel yet often intersecting paths. The first concerns near-integrable systems with a high degree of spatial symmetry, such as coupled-oscillator chains or particles in periodic lattices. In these cases, perturbative methods and the \textit{Kolmogorov–Arnold–Moser} (KAM) theorem provide a rigorous framework for understanding the persistence and breakdown of invariant tori \cite{Vladimir_I_Arnol'd_1963}. The second path addresses systems with quenched disorder, such as random potentials and rough energy landscapes, where the absence of symmetry shifts the focus toward statistical transport, diffusion, and localization phenomena \cite{PhysRevLett.49.509, PhysRevLett.67.3635}. A significant gap persists in the literature regarding potentials that occupy the middle ground: landscapes that are neither globally periodic nor statistically random, yet retain structured spatial organization. The present study belongs to this intermediate regime.

This article examines a specific, physically motivated example of how spatially localized, deterministically constructed, and pattern-inspired perturbations can lead to Hamiltonian instability and complex phase-space structures. To construct such a landscape, we draw inspiration from nodal pattern formation cause by vibration, which is known as cymatics. From Hooke's initial observations of nodal patterns in vibrating plates in $1680$ to Ernst Chladni's discovery of Chladni figures (nodal lines where the plate remains motionless) in the eighteenth century \cite{chladni1787entdeckungen, BB10515802}, we observe that these phenomena (later known as cymatics) represent some of the most visually captivating examples of different geometrical  arrangements/configurations emerging from vibrational energy. While studying rough surfaces and disordered potentials, it is observed that the landscapes are typically constructed through stochastic processes or periodic arrays. Such approaches, while mathematically tractable, discard the deterministic structure inherent in vibrational patterns. Distributing localized Gaussian perturbations along these nodal lines\textemdash what we term cymatics-inspired Gaussian landscape\textemdash creates a potential landscape that inherits this deterministic skeleton while acquiring the smooth, localized features amenable to dynamical analysis. Mathematically, a superposition of $N$ isotropic Gaussian functions has the following form
\begin{equation}\label{Eq1}
	V_{Gaussian}(\vec r) = V_{0} \sum_{i=1}^{N}e^{(\frac{-\abs{\vec r-\vec r_{i}}^{2}}{2\sigma^{2}})} 
\end{equation}
where  the set $\mathcal{C}=\{\vec r_{i}\}_{i=1}^{N}$ denotes sets of the bump centres, sampled from the nodal line geometry of a cymatic pattern. The perturbation remains tunable via three key parameters: the choice of  Chladni mode (governing the nodal geometry), amplitude of Gaussian bump ($V_{0}​$) and the spatial width of Gaussian bump ($\sigma$). Crucially, the underlying cymatic geometry imprints a long-range spatial correlation that is fundamentally different from both white noise and periodic functions. This offers a unique opportunity to explore the role of structured disorder/perturbation in dynamical evolution.

\begin{figure}[hbt!]
	\centering
	\fbox{\includegraphics[width=0.7\linewidth]{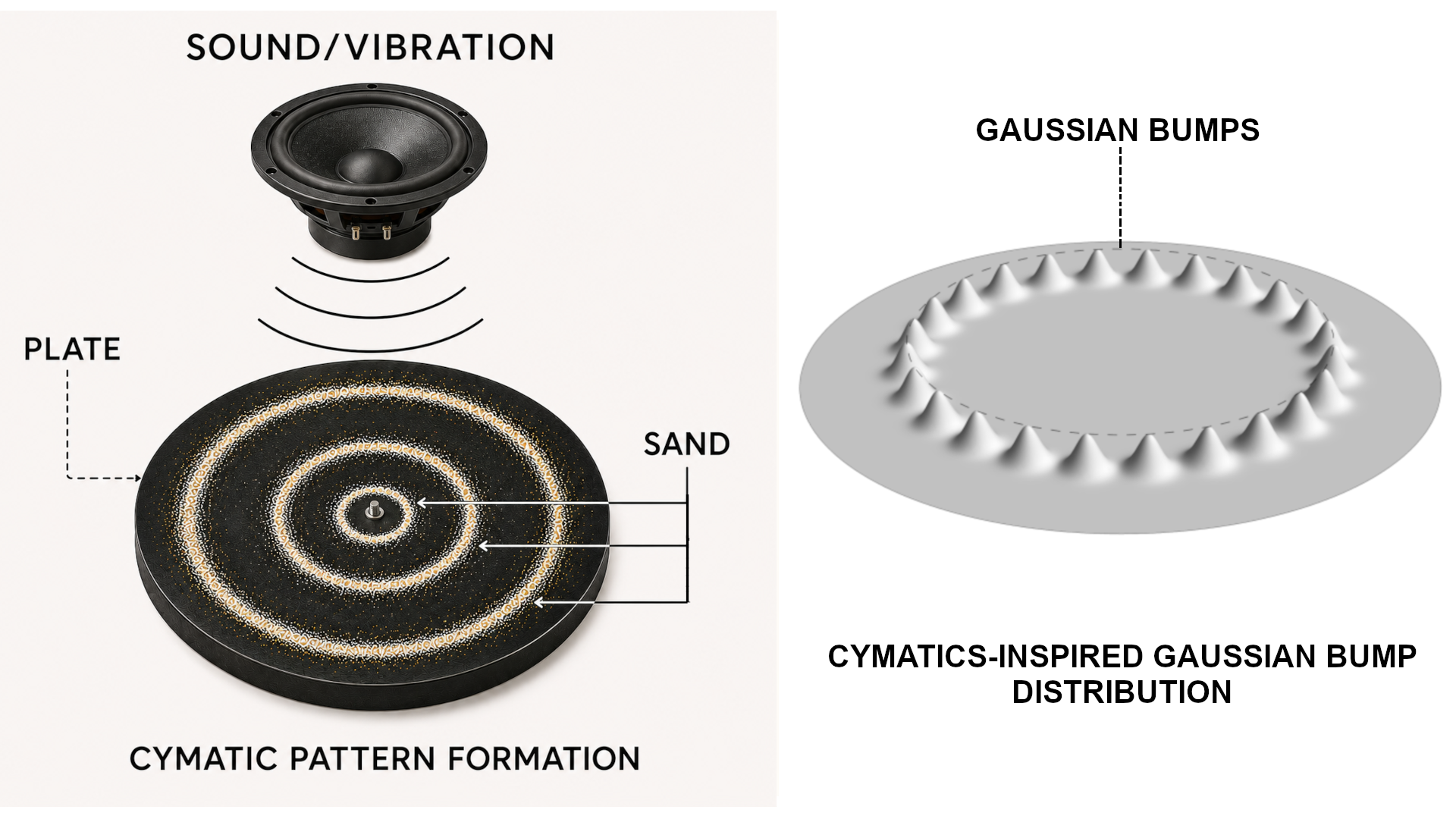}}
	\caption{Schematic diagram of cymatic pattern formation and correspondingly inspired Gaussian bump distributions. \textbf{Left:} An external acoustic excitation of a vibrating plate generates standing-wave modes (Chladni-type patterns) in a granular medium, causing sand to accumulate along nodal lines where the transverse displacement vanishes. \textbf{Right:} Idealized mapping of these nodal features, in which localized Gaussian bumps are arranged along a ring to emulate the symmetry and spatial organization of the experimentally observed nodal structures. This configuration serves as a controlled Hamiltonian perturbation, enabling systematic investigation of how patterned geometric localization influences phase-space structure, invariant torus stability, resonance overlap, and the emergence of chaotic dynamics.}
	\label{fig:cymaticsschematic}
\end{figure}

%LITERATURE
From extensive literature survey, we notice that several works have been published based on the Gaussian perturbed Harmonic oscillator. The harmonic oscillator with a Gaussian perturbation is an interesting model problem in its own right. Out of all these works, some publications demonstrate the fundamental analysis of the perturbed system \cite{bell1969gaussian, earl2008harmonic}. Some recent works present the study of the system from  quantum mechanical aspect, exploring the mechanism of scarring formation \cite{Luukko2016, Keski-Rahkonen_2019, PhysRevB.110.235420}. P. Amore \textit{et al.} \cite{Amore2024} and S. Fassari \textit{et al.} \cite{Fassari2020} studied the eigenvalues of the harmonic oscillator in the presence of a Gaussian perturbation. However, we observe that there is no analysis of such a Gaussian perturbed harmonic oscillator system based on chaos characterization from classical mechanics point of view, when Gaussian perturbed terms are arranged in certain cymatics-inspired patterns on the wall of the harmonic potential well. This observation motivates us to carry out a comprehensive study, which is the main objective of this chapter. 

Therefore, the central question this article explores is: How does the unique vibrational `skeleton' of a cymatic pattern affect the onset and characteristics of Hamiltonian chaos in a classical system, in comparison to purely periodic or purely random potentials?  To address this, we present a systematic numerical study of a particle moving through these cymatics-inspired Gaussian bump distributions. Our methodology  is anchored in  quantitative and qualitative diagnostic tools of nonlinear dynamics. We employ two principal tools: Lyapunov spectrum $\{\lambda_{i}\}$ and Poincaré map. Additionally, the diagnostic tools, namely phase portrait, time-series plot, power spectrum are also implemented to characterize the chaos.

%Outline
The remainder of this article is organized as follows. \cref{secII} establishes the mechanistic correspondence between cymatic pattern formation and Gaussian-structured Hamiltonian landscapes, providing the conceptual framework that motivates the geometric modelling approach. In \cref{secIII}, we introduce the  Hamiltonian formulation, parametric definitions and geometrical construction of several Gaussian-structured potential landscape models. In \cref{secIV}, different diagnostic tools are discussed.  \cref{secV} presents the results and discussion, where time-series plots, phase-space trajectories, Poincaré maps and Lyapunov spectra are analyzed to quantify the role of symmetry and spatial organization in the onset of chaos. Finally, \cref{secVI} summarizes the principal conclusions.

%%%%%%%%%%%%%%%%%%%%%%%%%%%%%%%%%%%%%%%%%%%
%%%%%%%%%%%%%%%%%%%

\section{\label{secII} Mechanistic correspondence between Cymatic pattern formation and Gaussian-structured Hamiltonian landscapes}
The analogy between cymatic pattern formation and Gaussian-structured Hamiltonian landscapes is grounded not merely in visual resemblance but in shared mechanisms of spatial organization under deterministic laws. In cymatics phenomena, boundary-constrained standing-wave modes generate nodal and anti-nodal regions through constructive and destructive interference, leading to the accumulation of granular material along dynamically stable nodal lines\textemdash regions of minimal transverse displacement. The resulting geometry is controlled by external parameters such as driving frequency and plate symmetry, which select and refine the vibrational patterns. In the Hamiltonian setting considered here, a deterministic arrangement of localized Gaussian perturbations generates a structured potential landscape whose extrema and saddle configurations are governed by spatial placement, amplitude, and width parameters. Just like vibrational frequency regulates nodal density in cymatics, the Gaussian amplitude and localization scale control resonance strength and spatial interaction in the potential well. In both systems, geometry encodes stability: nodal lines correspond to dynamically preferred regions in the wave field in case of cymatics phenomena, whereas structured and perturbed potential well landscapes organize the motion in phase space in case of Gaussian perturbed Hamiltonian systems. Thus, the Gaussian bump distribution serves as a dynamical analogue of cymatic spatial organization, translating interference-induced pattern formation into a controlled framework for studying  Hamiltonian  chaos.

While several choices for the ``bump'' function exist (e.g., Lorentzian, $ 1/(1+(r/a)^2)$; a finite-support bump, $ (1-(r/R)^2)^2$ for $r<R$; or a conical peak, $\max(0, 1 - r/R)$), the selection of the Gaussian functional form introduced in \cref{Eq1} offers clear analytical and dynamical advantages.  Among these functions, the most significant one is the Gaussian function because of its infinite differentiability, ensuring that the Hamiltonian and its associated vector field remain smooth everywhere in phase space. This regularity eliminates  scattering effects arising from non-smooth boundaries or discontinuous derivatives, allowing chaos to emerge solely from nonlinear resonance mechanisms. In addition, the Gaussian function decays faster than any algebraic power law, providing strong exponential localization.  Therefore, each Gaussian perturbation possesses a well-defined and tunable region of influence while avoiding long-range algebraic coupling and thereby allowing a controlled investigation of resonance interaction and torus breakdown. Moreover, the parameters governing amplitude, width and position of Gaussian bumps/dimples possess direct geometric interpretation, which enables systematic exploration of how spatial scale and symmetry influence the onset of chaos. For these reasons, the Gaussian profile constitutes a minimal, analytically robust, and geometrically interpretable choice for investigating patterned localization and the onset of Hamiltonian chaos.

Cymatics is the study of visualizing sound and vibration, typically using particles like sand on a vibrating plate. The resulting patterns form along the nodal lines (areas of minimum displacement and pressure variation), where the material collects. The areas of maximum pressure and displacement are antinodes,  where the material is repelled from. Different cymatic geometries (hexagonal, square, or star-shaped patterns) emerge from the interplay between boundary symmetry and harmonic excitation. For example, circular plates favour radially symmetric nodal rings, whereas polygonal boundaries imprint their symmetry on the nodal patterns; higher harmonics generate increasingly intricate nodal networks. Within the present Hamiltonian framework, localized Gaussian bumps serve as a mathematical replica for these nodal networks. By strategically placing bump centres to reflect specific harmonic symmetries and by fine-tuning their width to mimic modal sharpness, it is possible to create patterned arrangements that echo cymatic patterns. This approach offers complete control over the strength of disturbances and the extent of their interactions.
%%%%%%%%%%%%%%%%%%%%%%%%%%%
%%%%%%%%%%%%%%%%%%%%%%%%%%%%%%%%%
\section{\label{secIII} Models}
%%%%%%%%%%%%%%%%%%%%%%%%%%%%%
For serving this research, we consider a Hamiltonian that governs the motion of a test particle of mass $m ~(=1)$ within a $2D$ \textit{Harmonic Oscillator}(HO), which is given by:
\begin{equation}\label{chap7eq1}
	H(\vec p,\vec r)= \frac{{\vec p}^{2}}{2} + V_{\mathrm{eff}}(\vec r)
\end{equation}
where, momentum ($\vec p$) has two components i.e. $x-$component \& $y-$component and $V_{\mathrm{eff}}(\vec r)$ is the effective potential given as
\begin{equation}
V_{\mathrm{eff}}(\vec r)=V_{\mathrm{HO}}(\vec r)+ V_{\mathrm{pert}}(\vec r) + V_{\mathrm{noise}}(\vec r)
\end{equation}
	where
	\begin{eqnarray}
	V_{\mathrm{HO}}(\vec r)&=&\frac{\abs{\vec r}^{2}}{2} \qquad \text{where} \quad(\omega=1), \\
	V_{\mathrm{pert}}(\vec r) &=& V_{p}  e^{-(\frac{\abs{ \vec r}^{2}}{2\sigma^{2}})} ,  \\ \text{and}\quad 
	V_{\mathrm{noise}}(\vec r) &=& V_{n} \sum_{i}e^{(\frac{-\abs{\vec r-\vec r_{i}}^{2}}{2\sigma^{2}})} 
\end{eqnarray}

\begin{figure}[htbp!]
	\centering
	\includegraphics[width=\linewidth]{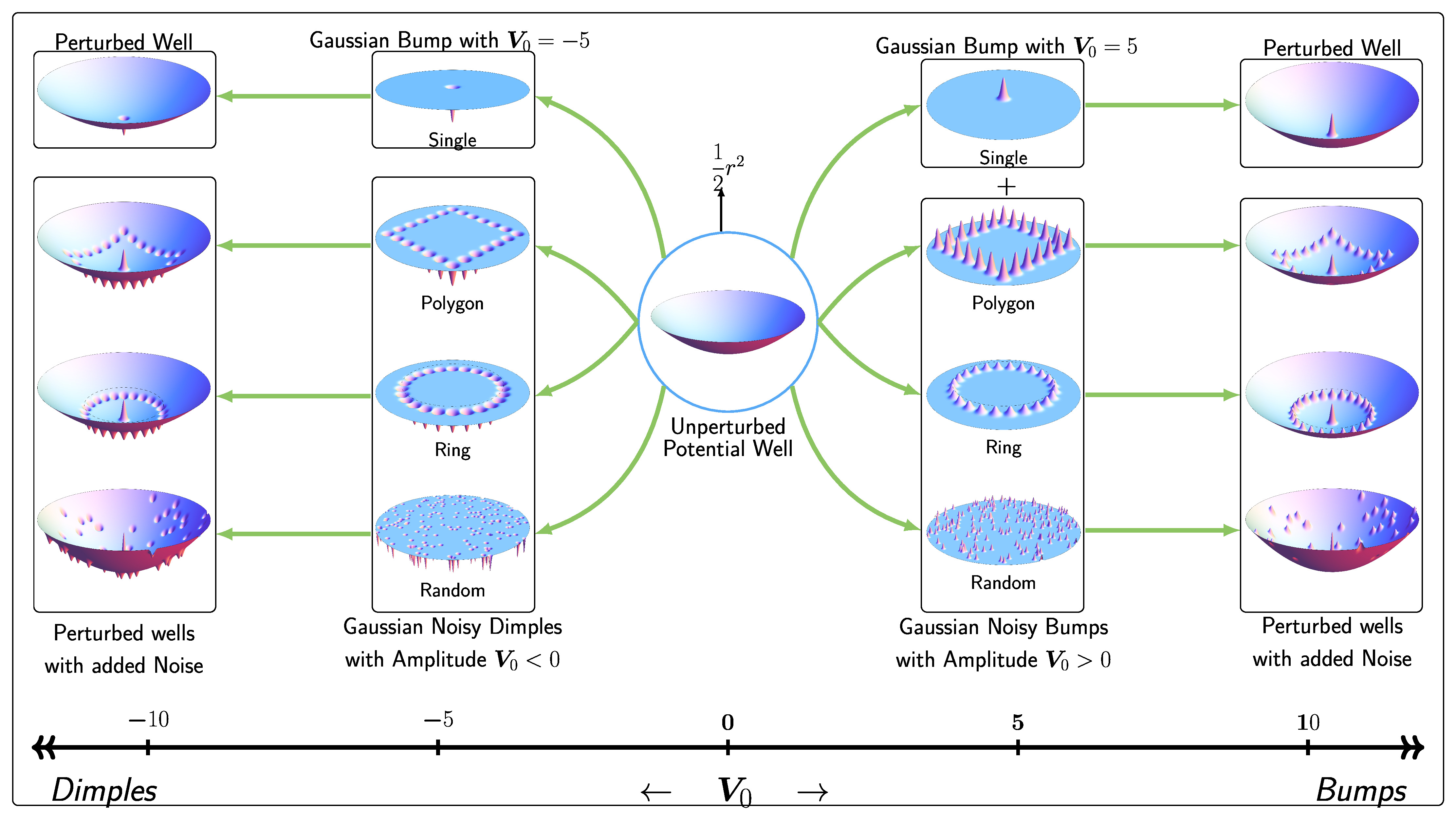}
	\caption{\label{schematic} Schematic illustration of the construction of a cymatics-inspired harmonic potential well $V_{\mathrm{HO}} (r ) = \frac{1}{2}r^{2}$. Perturbation terms with positive and negative amplitudes are named as ``bumps'' and  ``dimples'' respectively. Here, bumps/dimples are arranged in different geometrical shapes. For our study, we consider a fixed value for width of the bumps/dimples i.e.  $\sigma$=$\frac{0.235}{2 \sqrt{2 Log[2]}}$.}
\end{figure}

Equation (\ref{chap7eq1}) generates Hamilton's equations of motion: $\dot{r_{i}}=\{r_{i},H\}=\pdv{H}{p_{i}}$ and $\dot{p_{i}}=\{p_{i},H\}=-\pdv{H}{r_{i}}$. Here, we  assume $\omega=1$ without loss of generality. The system is autonomous and conservative, with total energy $\mathcal{E}$ as the only constant of motion, constraining trajectories to a three-dimensional energy hypersurface embedded in the four-dimensional phase space. The transition to chaos in this system is exquisitely sensitive to the functional form of $V_{G}(\vec r)$, which is divided into two parts: the term $V_{\mathrm{pert}}(\vec r)$ represents a controlled, deterministic primary Gaussian perturbation centred at origin (0,0) and the term $V_{\mathrm{noise}}(\vec r)$​ controls the distribution of localized Gaussian impurities centred at positions $\vec r_{i}$, representing weaker spatial inhomogeneities and these secondary perturbed terms (noise/impurities) are chosen randomly or in a geometric pattern. The parameters $V_p$ and $V_n$ control the amplitudes of the principal perturbation and impurity terms, respectively, and therefore quantify perturbation strength. Here, amplitudes $V_{n}$ are individually weaker than the perturbation amplitude $V_{p}$. Positive amplitudes correspond to  \textit{bumps}, while negative amplitudes generate  \textit{dimples} as shown in \cref{schematic}. For our study, we consider a fixed value for width of the bumps/dimples i.e.  $\sigma$=$\frac{0.235}{2 \sqrt{2 Log[2]}}=0.09979$.
%%%%%%%%%%%%%%%%%%%%%%%%%%%%%
%%%%%%%%%%%%%%%%%%%%%%%%%%%%%%%%%%%%%%%%%%%%%%%%%%%%%%%%%%%%%%%%

These structured perturbations generate patterned potential landscapes while preserving the conservative nature of the underlying Hamiltonian. To systematically investigate the role of spatial organization of the Gaussian perturbations (bumps and dimples) in the emergence of chaos, we analyze four distinct models, each containing total 25 number of Gaussian perturbed terms:
\begin{enumerate}[label={(\roman*)}]
	\item \textbf{Single:} A solitary Gaussian bump/dimple placed at the origin $(0,0)$ (\cref{contourplot}(b)) with finite $V_{p}$ and no additional impurities ($V_n = 0$). This configuration retains maximal radial symmetry and serves as the minimal perturbative reference.
	
	\item \textbf{Polygon:} A fixed central bump/dimple (finite $V_{p}$) accompanied by additional 24 bumps/dimples (finite $V_{n}$) arranged on the edges of a square (\cref{contourplot}(c)). Here, $V_{n} \leq V_{p}$. This arrangement introduces discrete rotational symmetry and controlled inter-perturbation coupling.
	
	%%%%%%%%%%%%%%%%%%%%%%%%%%%%%%%%%%%%%%%%%%%%%%%%%%%%%%%%%%%%%%%%%%%%%%%%%%%%%
\begin{figure}[hbt!]
	\centering
	\includegraphics[width=\linewidth]{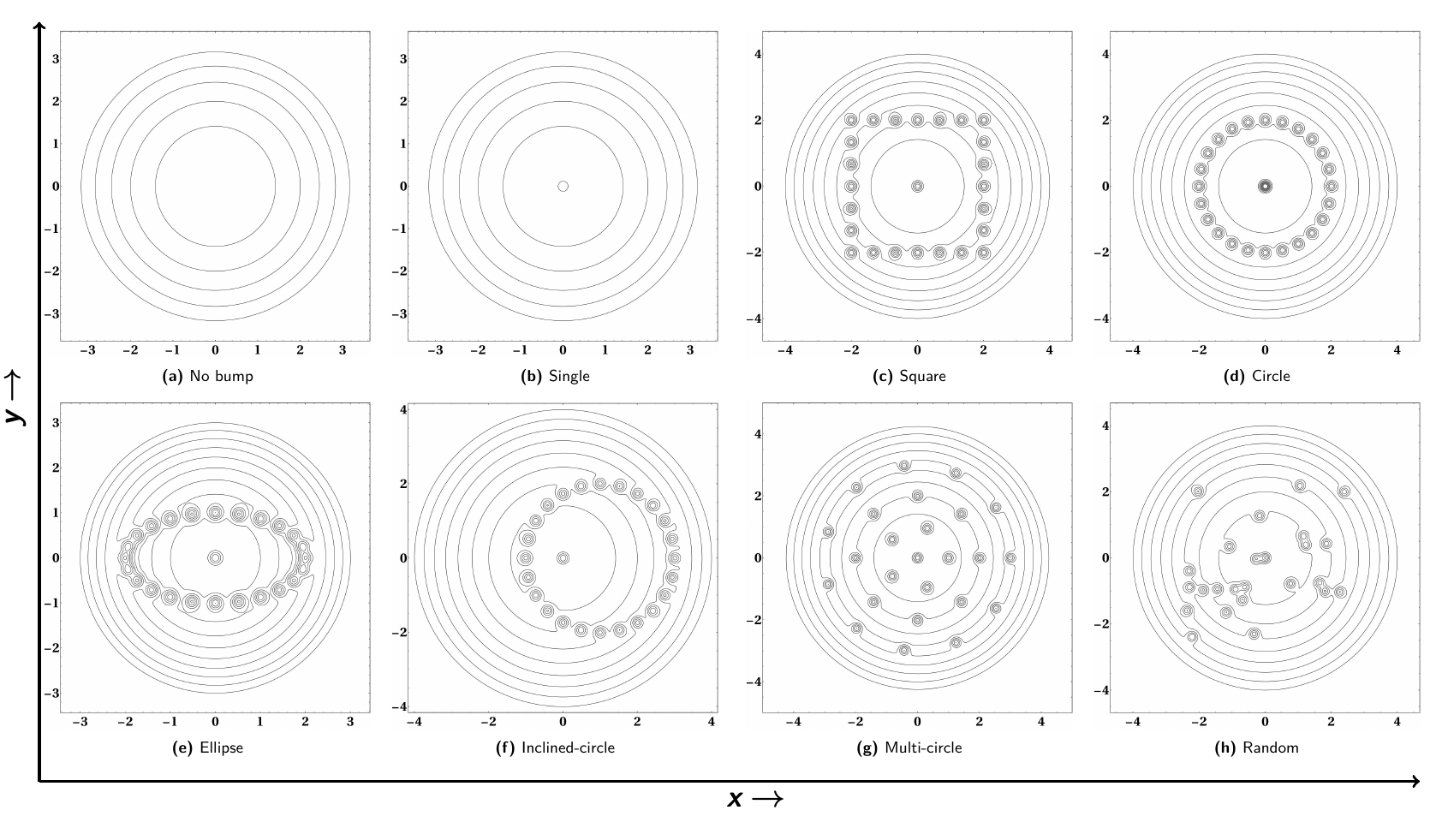}
	\caption{Contour plots illustrate the top-view spatial distribution of Gaussian bumps/dimples in the perturbed harmonic potential well, which are shown in configuration space. The sub-figure (a) indicates unperturbed harmonic well and the sub-figure (b) represents single central Gaussian perturbation. The sub-figures (c)-(h) depict a central bump/dimple supplemented by $24$ additional Gaussian perturbations arranged in square, circle, ellipse, inclined-circle, multi-circle (three concentric rings totalling $24$ bumps/dimples), and random configurations, respectively.}
	\label{contourplot}
\end{figure}
%%%%%%%%%%%%%%%%%%%%%%%%%%%%%%%%%%%%%%%%%%%%%%%%%%%%%%%%%%%%%%%%%%%%%%%%%%%

\item \textbf{Ring:} A fixed central bump/dimple (finite $V_{p}$) together with  peripheral distribution of 24 bumps/dimples (finite $V_{n}$, where $V_{n} \leq V_{p}$) along closed curves. The centres of these peripheral Gaussian terms are parametrized as:
\begin{equation}
	(x,~ y) = \big(a\cos(\frac{2 \pi k}{24})+r_{1},~ b\sin(\frac{2 \pi k}{24})+r_{2}\big)
\end{equation}
generating following ring-shape geometries, which are  respectively shown in \cref{contourplot}(d), (e), (f) \& (g):
\begin{itemize}
	\item \emph{Circle:} $a=b=2$, $r_{1}= r_{2}= 0$, $k= [0, 23] $ 
	\item \emph{Ellipse:} $a=2, b=1$, $r_{1}= r_{2}= 0$, $k= [0, 23]$
	\item \emph{Inclined-circle:} $a=b=2$, $r_{1}=1$, $r_{2}=0$, $k= [0, 23] $
	\item %\begin{flalign*}&
	\text{\emph{Multi-circle:}}
	$\begin{cases} 
		a=b=1,~ r_{1}= r_{2}= 0,~ k= [0, 4] : \text{$1^{st}$ circle with $5$ bumps} \\
		a=b=2,~ r_{1}= r_{2}= 0,~ k= [0, 7] : \text{$2^{nd}$ circle with $8$ bumps} \\
		a=b=3,~ r_{1}= r_{2}= 0,~ k= [0, 10] : \text{$3^{rd}$ circle with $11$ bumps} 
	\end{cases}$%& 
	%	\end{flalign*}
	\end{itemize}
	These configurations systematically vary symmetry, eccentricity, radial hierarchy, and inter-bump spacing.
	
	\item \textbf{Random:} A fixed central single bump/dimple accompanied by additional 24 bumps/dimples ($V_{n} \leq V_{p}$) placed at randomly distributed spatial locations, thereby breaking geometric symmetry and serving as a reference for disorder-induced effects (\cref{contourplot}(h)).
\end{enumerate}

Collectively, these models provide a structured progression from continuous symmetry (single), to discrete symmetry (polygonal and circular), to symmetry distortion (elliptical and shifted), and finally to symmetry-breaking randomness. Here, the secondary perturbation term $V_{noise}(\vec r)$ (noise/impurities)  play a crucial role to propose the model systems. This allows for a quantitative assessment of how geometric structure governs resonance interactions and chaotic transport in phase space. In other words, we can state that  we examine the effects of deformations (that arises on the wall of potential well due to Gaussian perturbations) on the development of chaotic dynamics in this study.

%%%%%%%%%%%%%%%%%%%%%%%%%%%%%%%%%%%
\section{\label{secIV}Chaos diagnostic tools}
To quantitatively characterise the transition from regular to chaotic motion in structured Gaussian landscapes, we employ a set of complementary chaos diagnostics. The emergence of Hamiltonian chaos in conservative systems manifests through several interconnected phenomena: the loss of predictability, the exponential divergence of nearby trajectories, the destruction of invariant tori, and the fractalisation of phase-space structures. No single diagnostic captures this complexity entirely. We therefore employ a hierarchy of tools\textemdash phase-space projections, Poincaré surfaces of section, Lyapunov spectra, and the Kaplan–Yorke dimension\textemdash to distinguish regular motion from weak chaos, quantify exponential sensitivity, and estimate the effective dimensionality of the explored invariant set. Together, these tools allow us to detect resonance overlap, invariant torus breakdown, and chaotic transport, thereby providing a coherent framework for assessing the dynamical consequences of spatially patterned perturbations. The following subsections outline several numerical tools that are tailored to our specific Hamiltonian landscapes.

%%%%%%%%%%%%%%%%%%%%%%%%%%%%%%%%%%%%%%%%%%%%%%%%%
\subsection{\label{secIVa}Time-series and phase-space}
We begin with direct inspection of phase-space trajectories and their associated time-series, which constitute the primary data set from which all subsequent diagnostics are derived. For a given initial condition, we numerically integrate Hamilton's equations of motion,
\begin{eqnarray}
	\dot{x}=\pdv{H}{p_{x}}=p_{x},&\quad& \dot{y}=\pdv{H}{p_{y}}=p_{y},\nonumber\\ \dot{p_{x}}=-\pdv{H}{x}=-\pdv{V_{\mathrm{eff}(x,y)}}{x},&\quad& \dot{p_{y}}=-\pdv{H}{y}=-\pdv{V_{\mathrm{eff}}(x,y)}{y}
\end{eqnarray}
using an $8^{th}-$order Runge-Kutta method with adaptive step-size control, ensuring energy conservation to within $10^{−8}$ relative error over the integration time $t=10^6$ (in natural units). initial conditions are sampled systematically across the accessible phase space and trajectories are recorded at uniform intervals $\Delta t$, yielding the scalar time-series $\{x(t), y(t), p_x(t), p_y(t)\}$.

Direct inspection of the time-series can provide preliminary insights. Regular, quasi-periodic motion manifests as amplitude-modulated oscillatory signals with discrete frequency spectra, while chaotic motion produces irregular, noise-like time-series with broadband power spectra. The autocorrelation function $C(\tau)=\frac{\expval{f(t)f(t+\tau)}}{\expval{f^{2}}}$, another qualitative discriminator, decays rapidly for chaotic dynamics, reflecting the loss of memory inherent in sensitive dependence on initial conditions. While these observations are qualitative, they provide an immediate indication of the degree of dynamical instability.

A complementary geometric insight into the structure of motion is obtained from phase-space trajectories. For our two-dimensional configuration space, the phase space is four-dimensional: $\{x,y,p_{x}​,p_{y}\}$. Although complete $4D$ visualisation is not possible, informative projections\textemdash such as $(x,p_x)$ or $(y,p_y)$ planes\textemdash capture essential structural features. 

Nevertheless, time-series inspection and phase-space projections alone are insufficient to definitively classify dynamics, particularly in mixed regimes where regular islands coexist with stochastic layers. This limitation necessitates more refined topological diagnostics. We therefore proceed to construct the Poincaré map, which reduces dimensionality while preserving the essential geometric structure of the Hamiltonian flow.

%%%%%%%%%%%%%%%%%%%%%%%%%%%%%%%%%%
\subsection{\label{secIVb}Poincaré map}
Because the full four-dimensional phase space is not directly visualisable, we construct a Poincaré map \cite{HOLMES1990137, shahhosseini2023poincare} that captures the recurrence properties of trajectories while preserving the essential features of the underlying symplectic dynamics. By intersecting trajectories with a suitably chosen two-dimensional hypersurface transverse to the flow, we reduce the continuous-time dynamics to a discrete area-preserving map.

We construct the Poincaré map by sampling the phase space each time the trajectory crosses the plane $y=0$ with positive momentum $p_{y}>0$. This choice is natural, given the symmetry of our cymatic patterns, and ensures that each crossing is transverse to the flow. The intersection condition $y=0$ reduces the four-dimensional state to a point ($x, p_{x}​$) at the moment of crossing, since $y$ is fixed and $p_{y}$​ is determined by energy conservation: $p_{y}​=\sqrt{2E​-2V(x,y=0)​-p_{x}^{2}}>0$. The resulting Poincaré map $P:(x_{n},p_{x,n})\mapsto(x_{n+1},p_{x,n+1})$ is area-preserving, reflecting the symplectic nature of the underlying Hamiltonian flow.

In integrable or near-integrable regimes, trajectories intersect the section along smooth invariant curves corresponding to quasiperiodic motion on invariant tori. Periodic orbits appear as isolated fixed points surrounded by chains of resonance islands. In contrast, chaotic trajectories generate scattered points that densely fill two-dimensional regions of the section, forming what is commonly termed a `chaotic sea'. As the perturbation amplitude increases, islands of regularity shrink, and the chaotic sea expands. The Poincaré map, therefore, provides a clear visual criterion for the onset of global chaos and transport across formerly invariant barriers.

By examining how the Poincaré section evolves with increasing energy or varying bump parameters, we directly observe the progressive destruction of KAM tori and the growth of chaotic regions\textemdash a visual narrative of the transition to global chaos. Because our models vary in geometric symmetry (single, polygonal, ring, random), the Poincaré map provides a direct visual measure of how spatial organisation influences resonance geometry.

%%%%%%%%%%%%%%%%%
\subsection{\label{secIVc}Lyapunov spectrum: $\{\lambda_{i}\}$}
While global  diagnostic tools reveal structural changes, quantitative characterisation of chaos requires measuring exponential sensitivity to ICs. This property is quantified by the Lyapunov spectrum ${\lambda_i}$ \cite{doi:10.1142/S021797929100064X, 2009, christiansen1997computing}. They measure of the average rate of divergence (or convergence) of nearby trajectories in phase-space of a dynamical system, thereby providing an objective, metric quantification of dynamical instability. They are pivotal in understanding the stability and chaotic behaviour of a system. Hamiltonian systems, characterised by conserved energy and phase-space volume, exhibit unique properties concerning Lyapunov exponents.

For a $d-$dimensional continuous-time dynamical system, there are $d$ exponents, typically ordered as $\lambda_{1}\geq \lambda_{2}\geq \dots\geq \lambda_{d}$​. While the full spectrum encodes the full hierarchy of stretching and contraction rates, the maximal Lyapunov exponent $\lambda_{1}=\lambda_{\max}$ is the most widely used diagnostic of chaos, defined as
\begin{equation}
	\lambda_{\max}=\lim\limits_{t\rightarrow0}~\lim\limits_{\Vert \delta (0)\Vert \rightarrow0}\frac{1}{t}\frac{\Vert \delta (t)\Vert}{\Vert \delta (0)\Vert}
\end{equation}
where $\delta z(t)$ represents the evolution of an initial infinitesimal perturbation.

In Hamiltonian systems, or more generally in conservative dynamical systems, positive Lyapunov exponents identify regions where nonlinear resonances, separatrix splitting, and the destruction of invariant tori lead to exponential sensitivity to initial conditions. Unlike dissipative systems that evolve towards attractors, Hamiltonian chaos exhibits a mixed-phase-space geometry where chaotic seas coexist with regular islands, and Lyapunov exponents vary across these structures. Consequently, the Lyapunov spectrum offers a precise measure of local instability and serves as a fundamental diagnostic tool for characterising the degree and spatial distribution of chaos in conservative dynamics. Several key properties of Lyapunov exponents in Hamiltonian systems are:
\begin{enumerate}[label={(\roman*)}]
	\item \textbf{Additive inverse pairs:} The symplectic symmetry and phase-space volume preservation (Liouville's theorem) ensure that Lyapunov exponents occur in additive inverse pairs $(\lambda_{i}= -\lambda_{d-i+1})$.
	\item \textbf{Zero exponents:} At least two exponents are always zero. While one of them indicates no divergence in one direction, the other is associated with energy conservation on the constant-energy hypersurface.
	\item \textbf{Vanishing sum:} The sum of all exponents vanishes ($\Lambda=\sum_{i=1}^{d} \lambda_{i}=0 $), reflecting phase space volume conservation.
\end{enumerate}

In integrable Hamiltonian systems, all Lyapunov exponents vanish (or form zero pairs), consistent with regular, quasi-periodic motion confined to invariant tori. In contrast, when nonlinear resonances overlap and invariant tori are destroyed, one or more exponent pairs become nonzero, with $\lambda_{\max}>0$ signalling chaotic dynamics. Importantly, Hamiltonian chaos differs fundamentally from dissipative chaos: rather than converging to attractors, trajectories evolve on a mixed phase space where chaotic regions coexist with regular islands. Consequently, LEs may vary across different initial conditions, reflecting the underlying phase-space structure.

In this work, we compute the complete Lyapunov spectrum using standard Wolf algorithm. This yields not only a robust measure of chaos via $\lambda_{\max}$​ but also the full spectrum necessary for computing the Kaplan-Yorke dimension. By tracking $\lambda_{\max}$ as a function of perturbation amplitude and energy, we obtain a quantitative measure of how Gaussian curvature and localisation scale control nonlinear resonance coupling. The magnitude of $\lambda_{\max}$ further provides an estimate of the characteristic timescale over which predictability is lost.

%%%%%%%%%%%%%%%%%%%%%%%%%%%%%%%%%%
\subsection{\label{secIVd}Kaplan-Yorke dimension: $D_{\mathrm{KY}}$}
The Kaplan-Yorke dimension\cite{Evans2000,Chen2018,doi:10.1142/S0218127422502224, CHLOUVERAKIS2005156} ($D_{\mathrm{KY}}$), also known as the Lyapunov dimension, provides a measure of the fractal dimensionality of the strange attractor\textemdash or, in the case of Hamiltonian chaos, the fractal structure of the chaotic region in phase space. While Hamiltonian systems do not possess attractors in the strict sense (due to phase space volume conservation), the chaotic component of a mixed phase space can exhibit complex, fractal boundaries and a cantor-like structure. The $D_{\mathrm{KY}}$ provides a direct way to quantify this complexity from the Lyapunov spectrum.

For a given dynamical system having $n$ number of LEs, the $D_{KY}$ is defined as 
\begin{equation}\label{eq10}
	D_{\mathrm{KY}} = m + \frac{\sum_{i=1}^{m} \lambda_{i}}{| \lambda_{m+1}|}
\end{equation}
where, $m$ is the largest integer for which the sum of the $m$ largest exponents is non-negative ( $\sum_{i=1}^{m} \lambda_{i} \geq 0$) and $\lambda_{m+1}$ is first LE for which  $\sum_{i=1}^{m+1} \lambda_{i} <0 $.

In practice, $D_{\mathrm{KY}}$​ serves as a compact, quantitative summary of the Lyapunov spectrum. In other words, we can state that  $D_{\mathrm{KY}}$ is a powerful geometric indicator that condenses the entire multidimensional Lyapunov spectrum into a single, intuitive metric. A value of $D_{\mathrm{KY}}$ close to the phase space dimension indicates that chaos occupies the most of the available volume of phase space (or almost the entire phase space). By computing ​$D_{\mathrm{KY}}$​ as a function of energy and bump parameters, we enhance our understanding of how Gaussian perturbations in the potential landscape influence both the onset of chaos and the fractal geometry of chaotic motion.

In  Hamiltonian systems, the Lyapunov spectrum satisfies $\lambda_{1}​>0, \lambda_{2}​=0, \lambda_{3}​=0, \lambda_{4}=-\lambda_{1}​$ for chaotic trajectories. The sum $\lambda_{1}+\lambda_{2}=\lambda_{1}>0$, while $\lambda_{1}+\lambda_{2}+\lambda_{3}=\lambda_{1}>0$, and $\lambda_{1}+\lambda_{2}+\lambda_{3}+\lambda_{4}=0$. In our case also, $\sum_{i=1}^{3} \lambda_{i}>0$ and $\sum_{i=1}^{4} \lambda_{i}=0$.  Thus, $m=3$ (since the sum of the first three is positive), and the Kaplan-Yorke dimension becomes:
\begin{equation}
	D_{\mathrm{KY}}  =3 + \frac{\sum_{i=1}^{3} \lambda_{i}}{\abs{\lambda_{3+1}}}=4
\end{equation}

%%%%%%%%%%%%%

%%%%%%%%%%%%%%%%%%%%%%%%%%%%%%%%%%%%%%%%%%%%%%%%%

\section{\label{secV}Result and Discussion}
In this section, we present the results of our classical dynamical studies on the cymatics-inspired Gaussian bump/dimple landscapes. The analysis combines the above said tools to quantify stability, resonance overlap, and chaotic transport.

%%%%%%%%%%%%%%%%%%%%%%%%%%%%%%%%%%%%%%%%%%%%%%%%%%%%%%%%%%%%%%
\subsection{\label{secVa}Single Bump Dynamics}
To isolate the role of cymatic-inspired noise, we begin with the minimal configuration: a two-dimensional harmonic oscillator perturbed by a single Gaussian bump, retaining $V_{\textrm{noise}}(\vec r)=0$. This configuration isolates the intrinsic effect of a smooth, localized deformation on an otherwise integrable system. The corresponding Hamiltonian is obtained from \cref{chap7eq1} by including only the $V_{\textrm{pert}}(\vec r)$ term. The transition from regular to chaotic dynamics is  primarily governed by two parameters\textemdash the perturbation amplitude $V_{p}$ and the spatial width $\sigma$.

\begin{figure}[hbt!]%[htbp!]
	\centering % <-- added
\includegraphics[width=\linewidth]{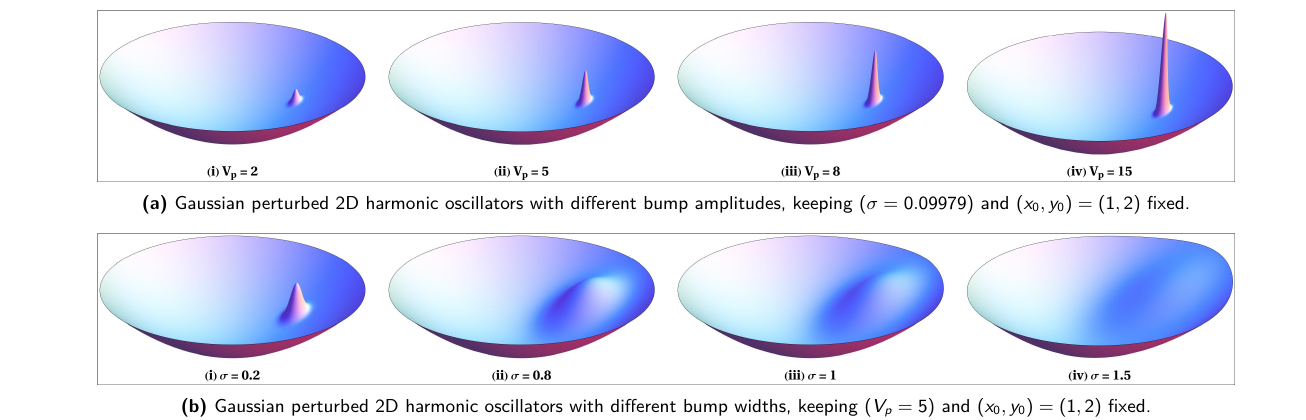}
	\caption{Single bump landscape models}
	\label{singlebump}
\end{figure}

To substantiate the scaling behaviour, we consider two complementary cases:
\begin{enumerate}[label={(\roman*)}]
	\item varying $V_{p}$ while keeping $\sigma$ and the bump centre fixed,
	\item varying $\sigma$ while keeping $V_{p}$ and the bump centre fixed, 
\end{enumerate}
as illustrated in \cref{singlebump}(a) and \cref{singlebump}(b), respectively.

%%%%%%%%%%%%%%%%%%%%%%%%%%%%%%%%%%%%%%

{	\renewcommand{\arraystretch}{1.5}
	\begin{table*}[hbt!]
		\caption{\label{table1} Lyapunov exponents,  their sum and Kaplan-Yorke dimension ($D_{KY}$) of a $2D$ harmonic potential well perturbed by a single Gaussian bump for different bump amplitudes ($V_{p}$) and different bump widths ($\sigma$). Here, position of the single bump is fixed at $(x_{0},y_{0})= (1,2)$ and all the Lyapunov exponents are calculated for a particular initial condition $(x(0),y(0), p_{x}(0),p_{y}(0))$. }
		%	\begin{ruledabular}
			
			\begin{minipage}[t]{0.48\linewidth}
				\centering
				\textbf{(a) Case-i (constant $\sigma$) \label{table1a}}
				\resizebox{\linewidth}{!}{
					
					\begin{tabular}{c c c c}
						\toprule[0.07cm]
						$V_{p}$ & LEs: $(\lambda_1,\lambda_2,\lambda_3,\lambda_4)$ & $\Lambda$ & $D_{KY}$ \\
						\midrule
						2.0  & (0.0064,0.0016,-0.0016,-0.0064) & $\sim0$ & 4 \\
						5.0  & (0.0113,0.0017,-0.0017,-0.0114) & $\sim0$ & 4 \\
						8.0  & (0.0138,0.0018,-0.0017,-0.0138) & $\sim0$ & 4 \\
						15.0 & (0.0169,0.0019,-0.0018,-0.0169) & $\sim0$ & 4 \\
						\bottomrule[0.07cm]
						\multicolumn{4}{@{}l}{\footnotesize $\sigma=0.09979$\qquad $\Lambda=\sum_{i=1}^{4}\lambda_{i}$}\\
					\end{tabular}
				}
			\end{minipage}
			\hfill
			\begin{minipage}[t]{0.48\linewidth}
				\centering
				\textbf{(b) Case-ii (constant $V_{p}$) \label{table1b}}
				\resizebox{\linewidth}{!}{
					
					\begin{tabular}{c c c c}
						\toprule[0.07cm]
						$\sigma$ & LEs: $(\lambda_1,\lambda_2,\lambda_3,\lambda_4)$ & $\Lambda$ & $D_{KY}$ \\
						\midrule
						0.2 & (0.0222,0.0015,-0.0027,-0.0210) & $\sim0$ & 4 \\
						0.8 & (0.0198,0.0066,-0.0069,-0.0195) & $\sim0$ & 4 \\
						1.0 & (0.0155,0.0057,-0.0088,-0.0124) & $\sim0$ & 4 \\
						1.5 & (0.0118,0.0043,-0.006,-0.0100) & $\sim0$ & 4 \\
						\bottomrule[0.07cm]
						\multicolumn{4}{@{}l}{\footnotesize $V_{p}=5.0$}\\
					\end{tabular}
				}
			\end{minipage}
		\end{table*}
	}

	%%%%%%%%%%%%%%%%%%%%%%%%%%%%%%%%%%%%%%%%%%%%%%%%%%%%%
	\begin{figure}[htbp!]
		\centering % <-- added
	\includegraphics[width=\linewidth]{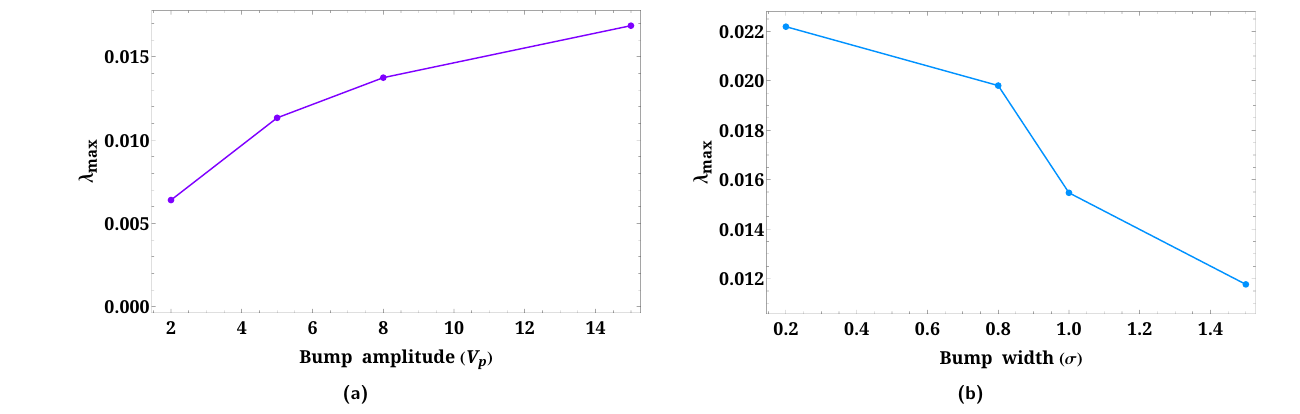}
		
		\caption{It shows that the largest Lyapunov exponent of a Gaussian perturbed harmonic potential well (a) increases with increase in bump amplitude ($\lambda_{max} \propto V_{p}$) and (b) decreases with increase in bump width ($\lambda_{max} \propto \frac{1}{\sigma}$) . The data listed in \cref{table1} is used to compute this plot.}
		\label{fig4}
	\end{figure}
	%%%%%%%%%%%%%%%%%%%%%%%%%%
	%%%%%%%%%%%%%%%%%%%%%%%%%%%%%%%%%%%%%%%%%%%%%%%%%%%%%
	Based on the data presented in \cref{table1}, we find that the two-dimensional harmonic potential well, perturbed by a single bump, exhibits chaotic motion regardless of the bump amplitude and width.  However, the level of chaos in the system increases with higher bump amplitudes and decreases with wider bump widths. This behaviour corresponds to the changes in the largest Lyapunov exponent ($\lambda_{1}$), as illustrated in \cref{fig4}. Each of these cases represents a conservative chaotic system, as indicated by the sum of Lyapunov exponents being zero, with corresponding $D_{KY}$ found to be an integer $4$, equal to the full phase-space dimension.

	A central observation is that chaoticity increases monotonically with bump amplitude ($V_{p}$) and decreases with bump width ($\sigma$) in a Gaussian-perturbed harmonic potential well. Physically, increasing $V_{p}$ enhances local curvature and strengthens nonlinear coupling between the unperturbed oscillator modes, thereby accelerating resonance overlap\textemdash a well-established route to global chaos in near-integrable Hamiltonian systems (see, e.g., Chirikov resonance-overlap criterion). Conversely, increasing $\sigma$ spatially broadens the perturbation, reducing local gradients and weakening effective mode coupling, which delays torus destruction and preserves regular motion over larger energy intervals.
	
	%%%%%%%%%%%%%%%%%%%%%%%
	%%%%%%%%%%%%%%%%%%%%%%%%%%%%%%%%%%%%%%%%%%%%%%%%%%
	%%%%%%%%%
	\subsection{\label{secVb} Cymatics-inspired Dynamics}

	In all models, the nonlinear deformation of the harmonic potential is introduced through Gaussian terms representing a single central perturbation and multiple surrounding impurities/noise. The central perturbation represents a dominant, structured deformation located at the origin with amplitude $V_p$. In contrast, the surrounding Gaussian terms represent secondary localized inhomogeneities that collectively modify the global topology of the potential landscape and it is characterized by amplitude $V_n$ and $V_n\leq V_p$ always.
	
	The amplitude of the central perturbation is fixed at $V_p = 5$, thereby establishing a consistent reference curvature and energy scale across all configurations. The amplitude $V_n$ of the surrounding Gaussian terms is then varied systematically and treated as a control parameter to quantify the onset and growth of chaotic dynamics. This separation of roles allows us to distinguish between the influence of a strong primary deformation and the cumulative effect of distributed secondary perturbations. By varying $V_n$​ within $[-5,5]$, both bump and dimple deformations are explored, enabling symmetric investigation of repulsive and attractive local modifications of the potential well. 
	
	The contour plots shown in \cref{contourplot} illustrate the resulting deformation patterns of the Gaussian-perturbed harmonic well for each arrangement. Except for the single-bump case, the total number of peripheral Gaussian bumps (or dimples) is fixed at 24 across all arrangements. By holding the number of perturbations fixed, any observed differences in dynamical behaviour can be attributed primarily to their spatial distribution rather than to variation in perturbation density. Consequently, the influence of geometric factors, such as symmetry, eccentricity, angular spacing, and radial hierarchy, can be isolated and systematically examined. Since Hamiltonian chaos is highly sensitive to local curvature variations and nonlinear resonance coupling, distinct geometric layouts generate different patterns of frequency modulation and resonance overlap. These differences are expected to manifest quantitatively in the Lyapunov spectrum and related instability measures, thereby linking spatial organization directly to dynamical complexity. Let us discuss the diagnosis of chaos by implementing various diagnostic tools in the forthcoming subsections.
	%%%%%%%%%%%%%%%%%%%%%%%%%%%%
	\subsubsection{\label{secVb1i}Time-series:}
		%%%%%%%%%%%%%%%%%%%%%%%%%%%%%%%%%%
	%%%%%%%%%%%%%%%%%%%%%%%%%%%%

\begin{figure}[htb!]
		\centering
		\includegraphics[width=\linewidth]{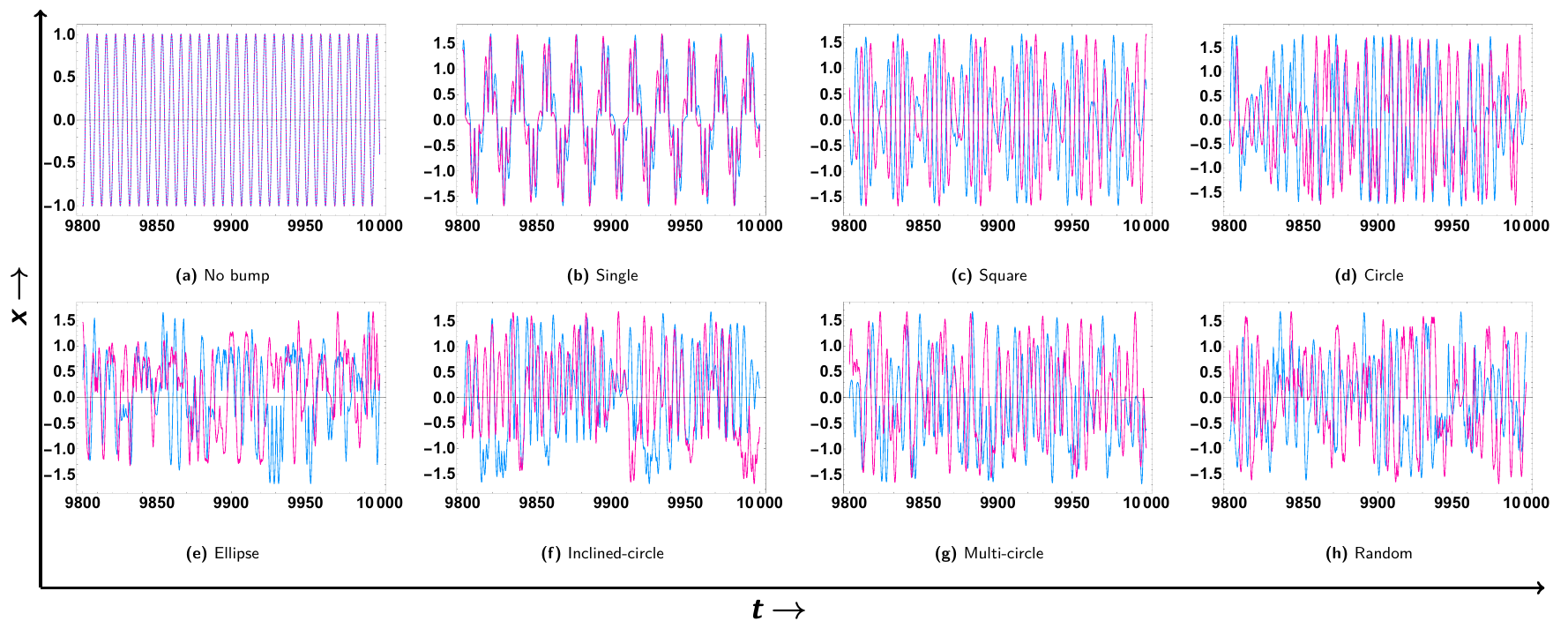}
		\caption{This figure represents the time-series plot of all model systems. The sub-figure (a) shows the periodic motion of the system, indicating the non-existence of chaos in a well-known unperturbed $2D$ harmonic potential well. All other seven sub-figures (b)-(h) represent sensitivity to initial conditions of each model system for two slightly different initial conditions. Here, we have considered that two initial conditions differ by $0.00001$ and two different colours in  sub-figures (b)-(h) illustrate the dynamics of corresponding perturbed model system along $x-$direction at two different initial conditions, indicating the existence of chaos due to emerging aperiodic behaviour in the system dynamics.}
		\label{timeseries}
	\end{figure}

	\cref{timeseries} presents a time-series of the $x-$coordinate for all configurations of model landscapes, providing a direct dynamical signature of regular versus chaotic motion. \cref{timeseries} (a), corresponding to the unperturbed harmonic oscillator, displays strictly periodic oscillations with constant amplitude and frequency, reflecting integrable dynamics and the absence of nonlinear coupling. The waveform remains perfectly regular over long  times, which is consistent with motion confined to invariant tori.
	
	In contrast, \cref{timeseries} (b)–(h), which include Gaussian perturbations, exhibit clear signatures of sensitive dependence on initial conditions. In each case, two trajectories are evolved from two nearly identical initial conditions that differ by $10^{-5}$ in phase-space coordinates.  Specifically, the initial conditions are chosen as $(x(0),y(0),p_{x}(0),p_{y}(0))=(0.1,0.2,1,1)$ and $(0.10001,0.20001,1.00001,1.00001)$, ensuring that the divergence of corresponding time-series trajectories and aperiodicity in each time-series arises from intrinsic dynamical instability.  Even though there is minute difference in two initial conditions,  their  initially overlapping trajectories gradually diverge for all perturbed configurations, leading to visibly distinct temporal evolutions after long times. The divergence is weak in the single-bump configuration (\cref{timeseries} (b)), where oscillations remain nearly regular with only slight phase drift, indicating weak chaos. However, when secondary Gaussian perturbations (noise) are introduced in addition to the central bump (\cref{timeseries} (c)-(h)), the time-series becomes visibly aperiodic. This progressive separation in trajectories due to sensitivity to initial conditions is a manifestation of exponential growth in trajectories, which is consistent with the numerically obtained positive largest Lyapunov exponent.

	Overall, the time-series analysis confirms that even small perturbative deformations in harmonic potential are sufficient to induce sensitive dependence on initial conditions. This observation reinforces the conclusion that the single-bump model represents the weakest chaotic regime, whereas multi-bump structured configurations exhibit enhanced dynamical instability and reduced long-term predictability, which is a characteristic of  chaos in conservative systems.

	%%%%%%%%%%%%%%%%%%%%%%%%%%%%%%%%%%%%%%%%%%%%
	\subsubsection{\label{secVb1ii}Phase-space trajectories:}
		\begin{figure}[hbt!]
		\centering
		\includegraphics[width=0.985\linewidth]{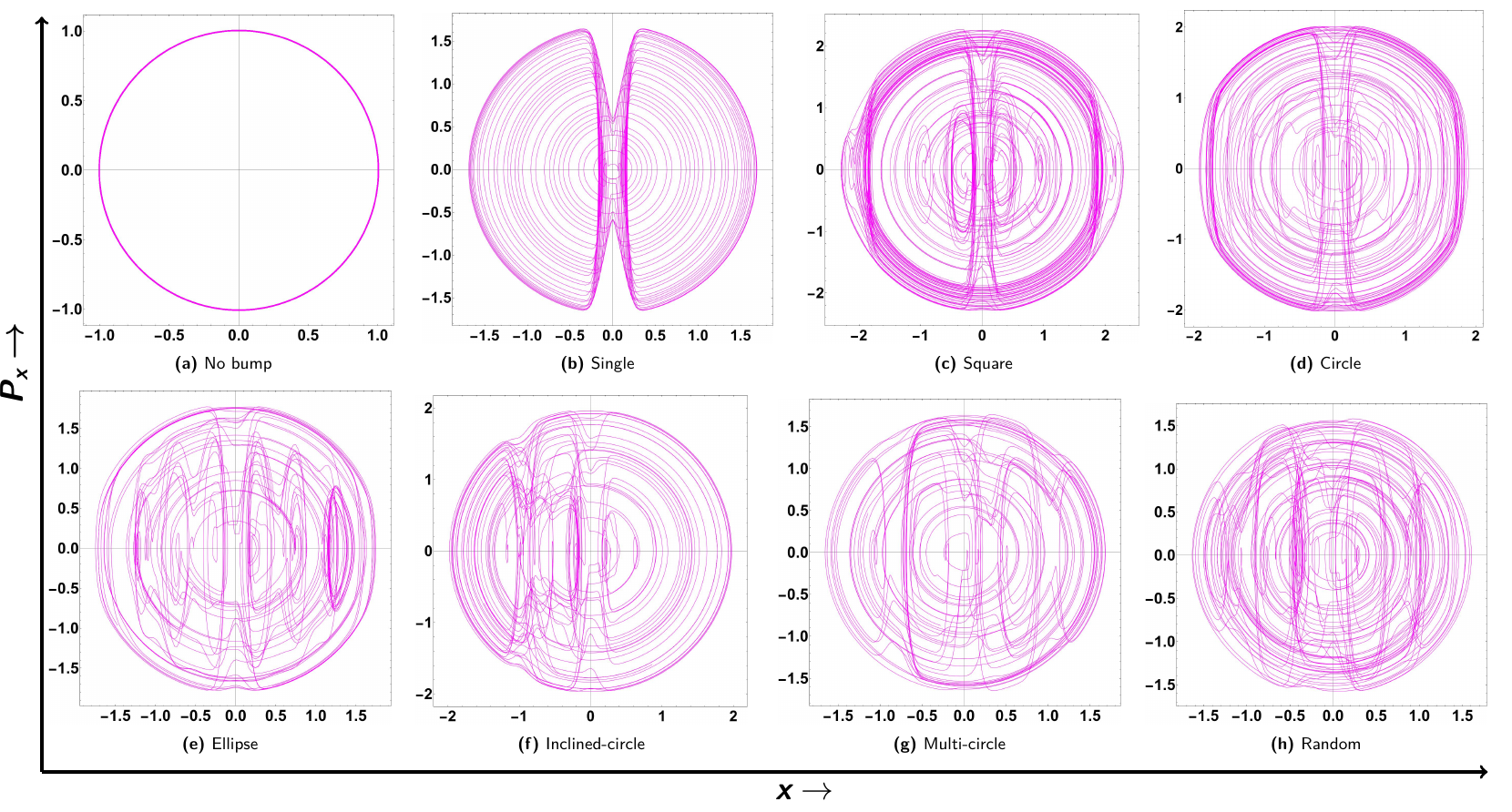}
		\caption{Phase-space trajectories  of the dynamical variables ($x(t), p_{x}(t)$) for different Gaussian perturbation landscapes: (a) unperturbed harmonic oscillator, (b) single, (c) square, (d) circle, (e) ellipse, (f) inclined-circle, (g) multi-circle, and (h) random arrangements of Gaussian bumps/dimples. The sub-figure (a) indicates regular periodic motion in case of unperturbed harmonic oscillator, whereas the sub-figures (b)-(h) illustrate irregular chaotic dynamics in Gaussian perturbed model systems.}
		\label{fig:ps}
	\end{figure}
	
	The phase-space trajectories for all model landscapes are shown in \cref{fig:ps}.  We know that  a pure (unperturbed) harmonic oscillator  behaves as an integrable regular system and therefore, the trajectories are smooth, closed curves on invariant tori, which is illustrated in \cref{fig:ps} (a). 
	
	Upon introducing Gaussian perturbations, regularity breaks down due to torus deformation, frequency modulation, and resonance overlap. As perturbation strength increases, resonance overlap leads to torus fragmentation, and chaotic trajectories begin to fill extended regions of the projected phase space densely. \cref{fig:ps} (b)-(h) indicate irregular and unpredictable dynamics of Gaussian perturbed model systems in $(x,p_x)$ phase space. It is observed that the motion in central single bump model system (\cref{fig:ps} (b)) is slightly irregular towards the centre of the landscape, where the deformation is formed due to presence of central Gaussian bump/dimple. Hence, there exist weak chaos in single bump system. However,  in case of other perturbed model systems, where multiple secondary Gaussian perturbations (or noise) are present along with a  primary central Gaussian perturbation (\cref{fig:ps} (c)-(h)), the trajectories are found to be highly irregular and unpredictable, indicating the existence of chaos. Henceforth, we conclude that the chaotic dynamics in case of considered perturbed harmonic oscillators is mainly noise-driven, though single bump model system (where noise is absent) exhibits weak chaotic dynamics.

	Therefore, the diagnosis of chaos in conservative Hamiltonian systems in phase space relies on identifying geometric signatures of instability within invariant manifolds and reduced regions of phase space. Regular motion in unperturbed harmonic oscillator is characterized by confinement to invariant tori, leading to periodic trajectories that remain ordered and structured over long times. In contrast, chaotic motion in Gaussian perturbed harmonic oscillators arises when nonlinear resonance interactions destroy these invariant tori, producing chaotic regions and allowing trajectories to explore extended regions of the energy manifold. This distinction becomes evident through projections such as Poincaré maps (demonstrated in forthcoming subsection), where regular dynamics appears as smooth invariant curves and resonance islands, while chaotic dynamics manifests as scattered point clouds filling finite areas.

%%%%%%%%%%%%%%%%%%%%%%%%%%%%%%%%%%%%%%%%%%
	\subsubsection{\label{secVb1iii}Power Spectrum:}
	We know that regular, quasi-periodic motion manifests as amplitude-modulated oscillatory signals with discrete frequency spectra, while chaotic motion produces irregular, noise-like broadband power spectra.
	
	\begin{figure}[htbp!]
		\centering
		\includegraphics[width=\linewidth]{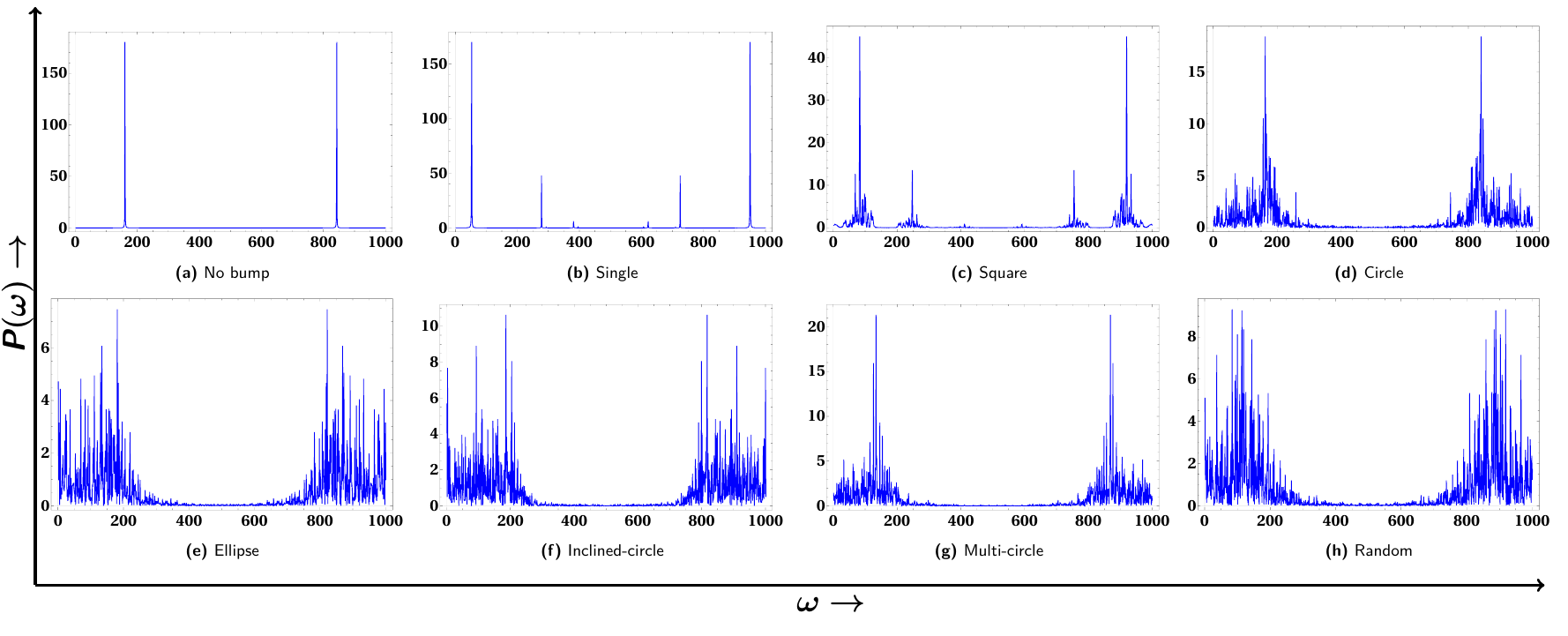}
		\caption{Power spectra for different Gaussian perturbation landscapes: (a) unperturbed harmonic oscillator, (b) single, (c) square, (d) circle, (e) ellipse, (f) inclined-circle, (g) multi-circle, and (h) random arrangements of Gaussian bumps/dimples.  The transition from discrete line spectra to broadband distributions demonstrates increasing resonance interaction and chaotic dynamics.}
		\label{fig:powerspec0allbump}
	\end{figure}

The qualitative distinction between regular and chaotic motion is remarkably obvious in the power spectrum shown in \cref{fig:powerspec0allbump}. In the unperturbed $2D$ harmonic oscillator (\cref{fig:powerspec0allbump} (a)), regular trajectories generate power spectra containing at most two incommensurate frequencies\footnote{Incommensurate frequencies means those frequencies whose ratios are irrational numbers and they do not share a common denominator or an integer multiple.}, which appear as sharp spikes.  The absence of additional harmonics\footnote{Harmonics are waves whose frequencies are exact integer multiples (e.g., 2x, 3x, 4x) of a primary base frequency, which is also known as the fundamental frequency.} in (\cref{fig:powerspec0allbump} (a))  confirms integrability and invariant-torus confinement. With the introduction of a single Gaussian bump (\cref{fig:powerspec0allbump} (b)), weak nonlinear coupling slightly modifies the natural frequencies. Additional spectral components appear due to nonlinear mixing, yet the spectrum remains predominantly discrete\textemdash indicating persistence of quasi-periodic dynamics with minor torus deformation.
	
	In contrast, for other perturbed model landscapes (\cref{fig:powerspec0allbump} (c)-(h)), $P(\omega)$ displays broadband with energy distributed continuously across a range of frequencies. This is a hallmark of developed Hamiltonian chaos, where nonlinear interactions generate effective frequency diffusion across a wide range of modes. The increasing geometric complexity of Gaussian bump arrangements enhances frequency coupling, accelerates resonance overlap, and promotes chaotic transport, as reflected directly in the power spectrum. In short, the power spectrum provides a frequency-domain diagnostic of torus breakdown and the emergence of Hamiltonian chaos in cymatics-inspired Gaussian landscapes.

 \cref{fig:powerspec0allbump} denotes  power spectra. For unperturbed oscillator (\cref{fig:powerspec0allbump} (a)), the corresponding blue colour plot is a line spectrum (absence of broadband distribution), indicating the ordered, predictable, and regular motion of the system. In contrast, for all other  Gaussian perturbed model systems,  the corresponding blue colour plots illustrated in \cref{fig:powerspec0allbump} (b)-(h)  reflect a broadband of continuous irregular noise-like spectrum, indicating the chaotic behaviour of the systems.

	%%%%%%%%%%%%%%%%%%%%%%%%%%%%
	\subsubsection{\label{secVb1iv}Auto-correlation function:}
		
	\begin{figure}[htbp!]
		\centering
		\includegraphics[width=\linewidth]{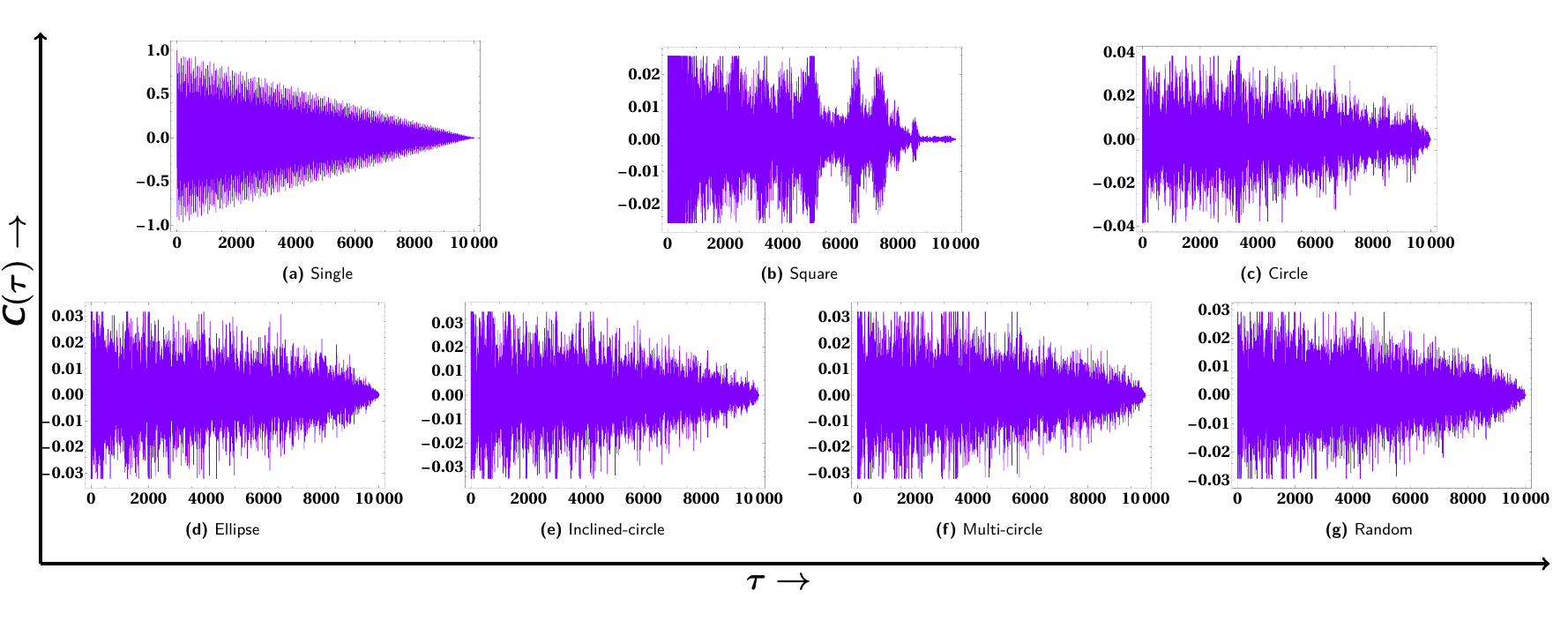}
		\caption{Auto-correlation function of the dynamical variable $x(t)$ for different Gaussian perturbation landscapes: (a) single, (b) square, (c) circle, (d) ellipse, (e) inclined-circle, (f) multi-circle, and (g) random arrangements. The correlation decay in sub-figures (a)-(g) indicate the existence of  chaotic dynamics.}
		\label{acf}
	\end{figure}
	
We know that the autocorrelation function is another qualitative diagnostic tool to confirm the existence of chaos in the system. The behaviour of long-time correlations is a metric for a 	trajectory’s stability and the deterioration of the correlation with time indicates whether 	a system is chaotic or not. Chaotic systems often have a quick correlation decay, but 	regular and mixed systems (coexistence of regular and chaotic motion) have no decay or a delayed decay, respectively.  \cref{acf}  illustrates the decay in correlations,thereby confirming  chaotic dynamics of all Gaussian perturbed model systems under study.

We observe that the time-series inspection, phase-space projections, power spectra and decay in correlations  alone are insufficient to definitively classify dynamics, particularly in mixed regimes where regular islands coexist. This limitation necessitates more refined  diagnostic tools. We therefore proceed to construct the Poincaré map, which reduces dimensionality while preserving the essential geometric structure of the Hamiltonian flow.

	%%%%%%%%%%%%%%%%%%%%%%%%%%%%
	%%%%%%%%%%%%%%%%%%%%%%%%%%%%%%%%%%%%%%%%%%%
	\subsubsection{\label{secVb2}Poincaré Map:}
	
	\begin{figure}[hbtp!]
		\centering % <-- added
			\includegraphics[width=18.0cm,height=23.25cm] {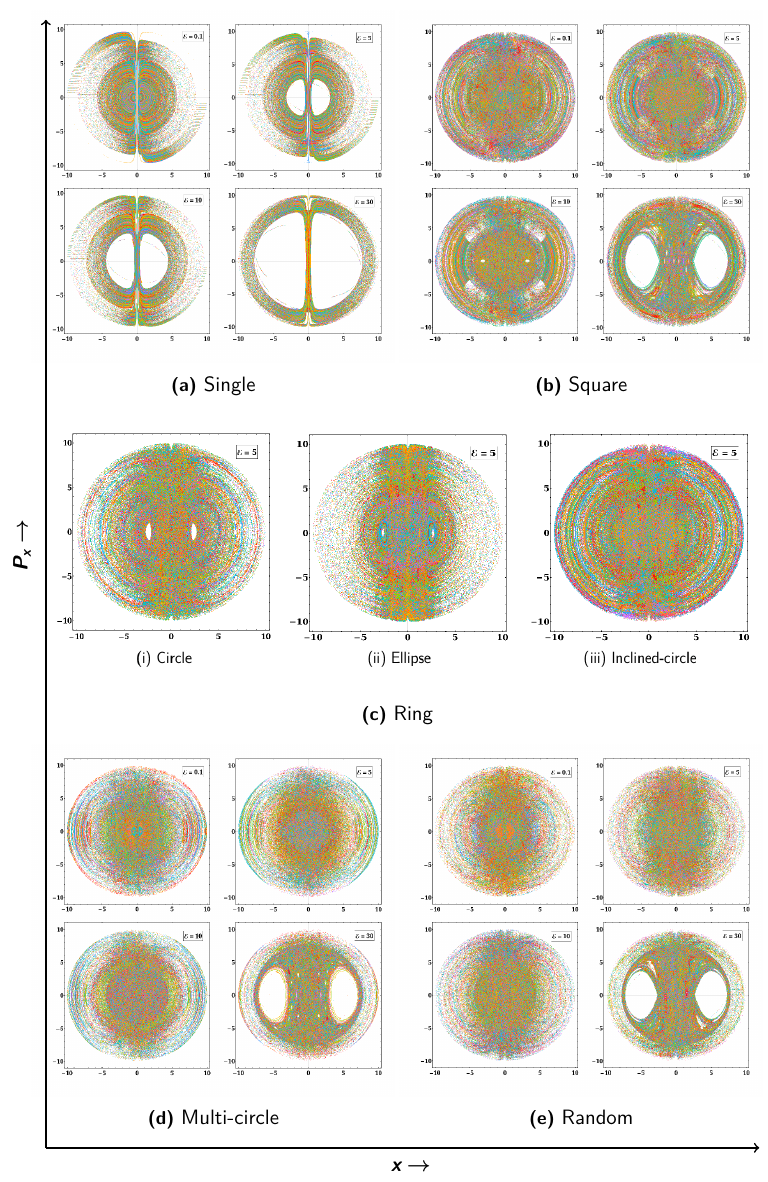}	
		\caption{Poincaré maps of each perturbed configurations at different energies.}
		\label{Poincaré-maps}
	\end{figure}
	
	\cref{Poincaré-maps} presents the Poincaré maps for the Gaussian-perturbed harmonic oscillator models at different total energies, which are computed for fixed amplitudes $V_{p}=5$ \& $V_{n}=2$ and for energy $\mathcal{E}\in\{0.1,5,10,30\}$. The corresponding Lyapunov exponents are tabulated in \cref{table2}.
	
	In the single-bump configuration (\cref{Poincaré-maps} (a)), strong chaotic dynamics is observed at low energy ($\mathcal{E}=0.1$), where scattered point clouds densely fill large portions of the cross-section plane. This observation aligns with the analysis of the Lyapunov exponent, which indicates maximal instability at low energy. As the energy increases to $\mathcal{E}=5,10,30$, progressively larger regions of the cross-section plane are replaced by smooth invariant curves and void-like areas that correspond to regular motion. Hence, for fixed perturbation amplitude, the influence of the localized Gaussian deformation diminishes monotonically with energy.

	A similar trend is also observed for other perturbed distributions (square: \cref{Poincaré-maps} (b), multi-circle: \cref{Poincaré-maps} (d) and random: \cref{Poincaré-maps} (e)). At low energy ($\mathcal{E}=0.1$), trajectories remain confined near the deformation centre, where local curvature gradients are strongest. The cross-section plane displays extensive chaotic regions with irregularly distributed points occupying substantial portions of the accessible region. Only small remnants of invariant curves survive, indicating significant resonance overlap. As the energy increases ($\mathcal{E}=5,10,30$), the phase-space portrait undergoes systematic reorganization. Regions of scattered points gradually contract, while smooth invariant curves and resonance islands become increasingly prominent. Although chaotic regions remain significant, the relative phase-space volume occupied by regular motion increases with increase in energy. This behaviour reflects the decreasing relative influence of the localized Gaussian deformation as trajectories explore regions farther from the perturbation core. In case of square landscape, it seems like the regions correspond to the corner of square shape bump distribution demonstrate low density region of unorganized points when energy increases, indicating less influence of perturbations on the dynamics of the system at these regions as compared to the other regions of phase space. In case of multi-circle landscape, the hierarchical organization of Gaussian perturbations  increases the complexity of the mixed-phase-space structure without altering the total perturbation strength. In case of random landscape, the absence of symmetry prevents systematic organization of resonances, often leading to more spatially dispersed but less geometrically structured chaotic regions.
	
	For the rounded/ring shaped model landscapes (circle, ellipse, and inclined-circle: \cref{Poincaré-maps} (c)), Poincaré-maps are computed at $\mathcal{E}=5$ and they indicate that densely populated chaotic regions coexist with resonance islands, demonstrating substantial nonlinear coupling induced by the distributed Gaussian perturbations. In the circular case, the nearly uniform angular spacing of Gaussian bumps leads to a more isotropic modulation of curvature of perturbed potential well, resulting in radially distributed points. In contrast, the elliptical configuration disrupts rotational symmetry and introduces anisotropy in curvature of the  perturbed  well, which creates asymmetric resonance structures and stability regions that depend on direction. The inclined-circular distribution of bumps (where the ring shaped arrangement shifts relative to the origin) breaks central symmetry while preserving the circular ordering among the perturbations. This results in a non-uniform distortion of invariant curves in the Poincaré map. Since the underlying Hamiltonian flow is canonical, the induced Poincaré map is area-preserving (symplectic). Consequently, Liouville's theorem guarantees conservation of phase-space volume, preventing the existence of dissipative attractors. Therefore, the observed complexity  arises purely from nonlinear resonance interactions and curvature-induced instability rather than from energy loss.

	As the total energy increases, trajectories explore regions farther from the localized perturbation, where the relative influence of the localized Gaussian perturbations diminishes compared to that of the quadratic harmonic confinement. Consequently, invariant tori re-emerge over substantial regions of phase space, and chaotic layers contract to narrower regions. This trend indicates a progressive restoration of near-integrable behaviour at higher energies for fixed perturbation amplitude. This energy-dependent suppression of chaotic behaviour is robust across all spatial arrangements (single, square, ring, and random). However, the detailed distributions of islands and chaotic regions depend sensitively on symmetry and radial hierarchy.

	Although Poincaré maps provide high-resolution geometric insight into invariant structures, they remain qualitative. For quantitative comparisons across models, we compute the full Lyapunov spectrum, providing a robust measure of dynamical instability.
	
%%%%%%%%%%%%%%%%%%%%%%%%%%%%%%%%%%%%%%%%%%%%%%%%%%%%%%%%%%%%%%%%%%%%%%%%%%%%%%
	\subsubsection{\label{secVb3}Lyapunov Exponents:}

	The Lyapunov exponents are computed using \texttt{Mathematica 14.2} by integrating both the equations of motion and the associated variational equations (means the dynamical equations of the system that are produced due to tiny changes in initial condition) and hence, the rate of divergences between nearby trajectories are measured. For each model, trajectories are evolved for $2\times10^{6}$ time steps with a fixed step size $\Delta t=10^{-4}$. Convergence of the exponents are monitored to ensure that the finite-time estimation  gives stable asymptotic (saturated) values of exponents. Enhancing the numerical precision beyond these parameters does not result in a significant change in the converged Lyapunov spectrum, demonstrating the robustness of the calculations.

	%%%%%%%%%%%%%%%%%%%%%%%%%%%%%%%%%%%%%%%%%%%%%%
	{	\renewcommand{\arraystretch}{1.5}
		\begin{table*}[htb!]
			\centering
			\footnotesize
			
			\caption{\label{table2}Lyapunov Exponents, their sum and Kaplan-Yorke dimension of each Gaussian perturbed model systems at different energies by considering fixed amplitudes $V_{p}=5$ \& $V_{n}=2$. It is observed that for all perturbed models, there exist two  positive Lyapunov exponents and  sum of all four Lyapunov exponents is found to be zero. These two observations confirm the existence of conservative  chaos.}
			
			\addtolength{\tabcolsep}{-0.6em}
			
			\begin{tabular}{l c c c c }
				\toprule[0.07cm]
				%	\hline\hline\\
				
				\bf{Models} &  \bf{Energy}  & \bf{LEs : ($\lambda_1, \lambda_2, \lambda_3, \lambda_4$)}& \bf{$\sum{\lambda_i}$} &  \bf{$D_{KY}$}  \\    
				
				\midrule
				
				\multirow{1}{*}{Single}& 
				$\begin{cases} 
					\mathcal{E}=0.1 \\
					\mathcal{E}=5 \\
					\mathcal{E}=10\\
					\mathcal{E}=30 
				\end{cases}$
				& 
				$\begin{cases} 
					$(0.02690,0.02116,-0.02135,-0.02673)$ \\
					$(0.01794,0.00119,-0.00527,-0.01388)$ \\
					$(0.01492,0.00020,-0.00441,-0.01071)$ \\
					$(0.00944,0.00012,-0.00445,-0.00488)$ 
				\end{cases}$
				& 
				$\begin{cases} 
					0  \\ 
					0  \\ 
					0 \\ 
					0 \\ 
				\end{cases}$
				& 
				$\begin{cases} 
					4\\ 
					4\\ 
					4\\ 
					4\\ 
				\end{cases}$
				\\ \addlinespace
				%	\rule{0pt}{8ex}
				
				\multirow{1}{*}{Square}& 
				$\begin{cases} 
					\mathcal{E}=0.1 \\
					\mathcal{E}=5 \\
					\mathcal{E}=10\\
					\mathcal{E}=30 
				\end{cases}$
				& 
				$\begin{cases} 
					$(0.02521,0.02043,-0.02046,-0.02521)$ \\
					$(6.02395,0.04970,-0.04971,-6.02394)$ \\
					$(2.94447,0.01548,-0.01548,-2.94447)$ \\
					$(1.61291,0.00056,-0.00056,-1.61291)$ 
				\end{cases}$
				& 
				$\begin{cases} 
					0  \\ 
					0  \\ 
					0 \\ 
					0 \\ 
				\end{cases}$
				& 
				$\begin{cases} 
					4\\ 
					4\\ 
					4\\ 
					4\\ 
				\end{cases}$
				\\ \addlinespace
				
				\multirow{1}{*}{Circle}& 
				$\begin{cases} 
					\mathcal{E}=0.1 \\
					\mathcal{E}=5 \\
					\mathcal{E}=10\\
					\mathcal{E}=30 
				\end{cases}$
				& 
				$\begin{cases} 
					$(0.02566,0.02325,-0.02327,-0.02566)$ \\
					$(5.26938,0.05803,-0.05803,-5.26938)$ \\
					$(2.55112,0.00555,-0.00555,-2.55112)$ \\
					$(0.34550,0.00139,-0.00225,-0.34465)$ 
				\end{cases}$
				& 
				$\begin{cases} 
					0  \\ 
					0  \\ 
					0 \\ 
					0 \\ 
				\end{cases}$
				& 
				$\begin{cases} 
					4\\ 
					4\\ 
					4\\ 
					4\\ 
				\end{cases}$
				\\ \addlinespace
				
				\multirow{1}{*}{Ellipse}& 
				$\begin{cases} 
					\mathcal{E}=0.1 \\
					\mathcal{E}=5 \\
					\mathcal{E}=10\\
					\mathcal{E}=30 
				\end{cases}$
				& 
				$\begin{cases} 
					$(0.02566,0.02325,-0.02325,-0.02566)$ \\
					$(0.93311,0.01007,-0.01026,-0.93293)$ \\
					$(0.47106,0.00214,-0.00229,-0.47091)$ \\
					$(0.09102,0.00081,-0.00141,-0.09042)$ 
				\end{cases}$
				& 
				$\begin{cases} 
					0  \\ 
					0  \\ 
					0 \\ 
					0 \\ 
				\end{cases}$
				& 
				$\begin{cases} 
					4\\ 
					4\\ 
					4\\ 
					4\\ 
				\end{cases}$
				\\ \addlinespace
				
				\multirow{1}{*}{Inclined-circle}& 
				$\begin{cases} 
					\mathcal{E}=0.1 \\
					\mathcal{E}=5 \\
					\mathcal{E}=10\\
					\mathcal{E}=30 
				\end{cases}$
				& 
				$\begin{cases} 
					$(0.02521,0.02043,-0.02046,-0.02521)$ \\
					$(0.95537,0.01787,-0.02440,-0.94884)$ \\
					$(0.68179,0.00109,-0.00103,-0.68187)$ \\
					$(0.17797,0.00128,-0.00208,-0.17718)$  
				\end{cases}$
				& 
				$\begin{cases} 
					0  \\ 
					0  \\ 
					0 \\ 
					0 \\ 
				\end{cases}$
				& 
				$\begin{cases} 
					4\\ 
					4\\ 
					4\\ 
					4\\ 
				\end{cases}$
				\\ \addlinespace
				
				\multirow{1}{*}{Multi-circle}& 
				$\begin{cases} 
					\mathcal{E}=0.1 \\
					\mathcal{E}=5 \\
					\mathcal{E}=10\\
					\mathcal{E}=30 
				\end{cases}$
				& 
				$\begin{cases} 
					$(0.02582,0.01963,-0.01809,-0.02637)$ \\
					$(0.90023,0.04321,-0.04325,-0.90019)$ \\
					$(0.55261,0.01494,-0.01494,-0.55263)$ \\
					$(0.15402,0.00204,-0.00197,-0.15409)$  
				\end{cases}$
				& 
				$\begin{cases} 
					0  \\ 
					0  \\ 
					0 \\ 
					0 \\ 
				\end{cases}$
				& 
				$\begin{cases} 
					4\\ 
					4\\ 
					4\\ 
					4\\ 
				\end{cases}$
				\\ \addlinespace
				
				\multirow{1}{*}{Random}& 
				$\begin{cases} 
					\mathcal{E}=0.1 \\
					\mathcal{E}=5 \\
					\mathcal{E}=10\\
					\mathcal{E}=30 
				\end{cases}$
				& 
				$\begin{cases} 
					$(0.02506,0.02367,-0.02429,-0.02447)$ \\
					$(0.65883,0.01622,-0.01405,-0.66101)$ \\
					$(0.42326,0.00418,-0.00334,-0.42411)$ \\
					$(0.26484,0.00110,-0.00119,-0.26475)$ 
				\end{cases}$
				& 
				$\begin{cases} 
					0  \\ 
					0  \\ 
					0 \\ 
					0 \\ 
				\end{cases}$
				& 
				$\begin{cases} 
					4\\ 
					4\\ 
					4\\ 
					4\\ 
				\end{cases}$
				\\
				
				%\hline\hline
				\bottomrule[0.07cm]
			\end{tabular}
		\end{table*}
	}

	%%%%%%%%%%%%%%%%%%%%%%%%%%%%%%%%%%%%%%%%%%%%%%%%%%%%%
	\begin{figure}[htb!]
		\centering % <-- added
		\includegraphics[width=0.7\linewidth]{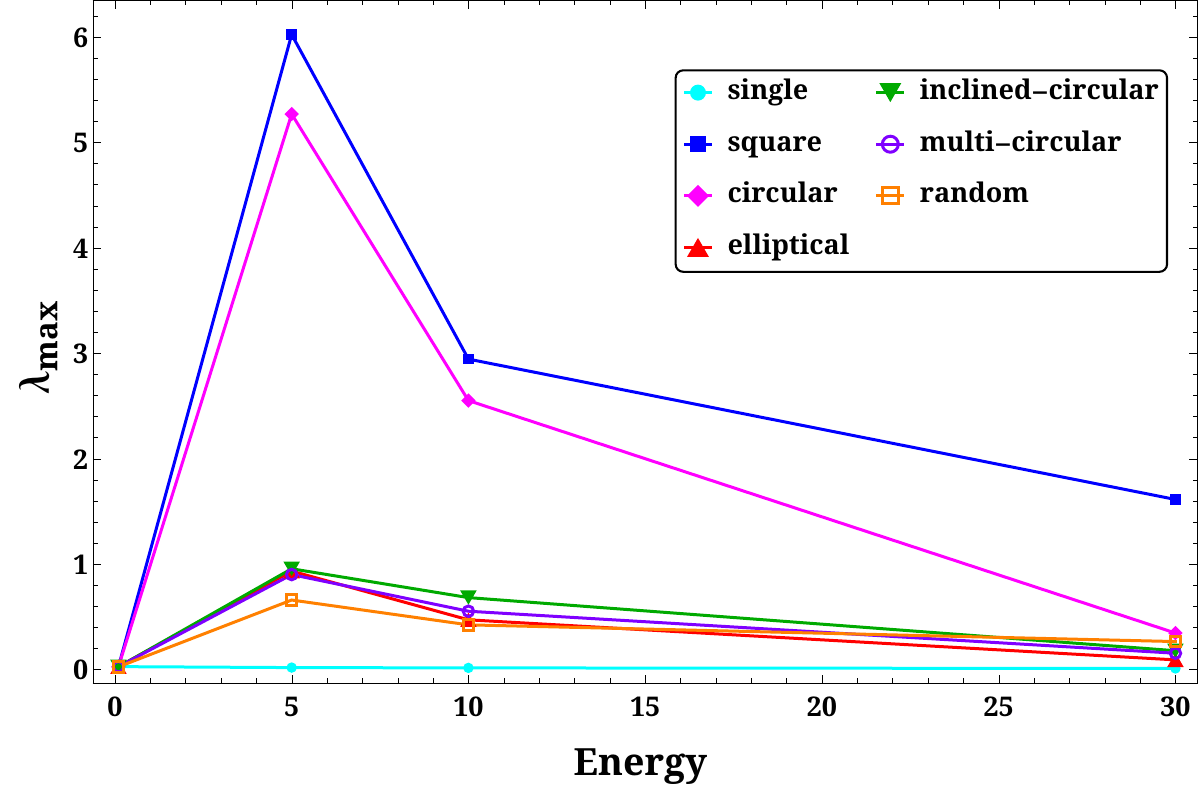}
		
		\caption{It shows that the largest Lyapunov exponent of any Gaussian perturbed harmonic potential well decreases with increase in energy for all model systems i.e. $\lambda_{max} \propto \frac{1}{\mathcal{E}}$. The data listed in \cref{table2} is used to compute this plot.}
		\label{le-vs-E-plot}
	\end{figure}
	%%%%%%%%%%%%%%%%%%%%%%%%%%%%%%%%%%%%%%%%%%%%%%%%%%
	%%%%%%%%%%%%%%%%%%
\begin{figure}[htb!]
	\centering % <-- added
	\includegraphics[width=0.98\linewidth]{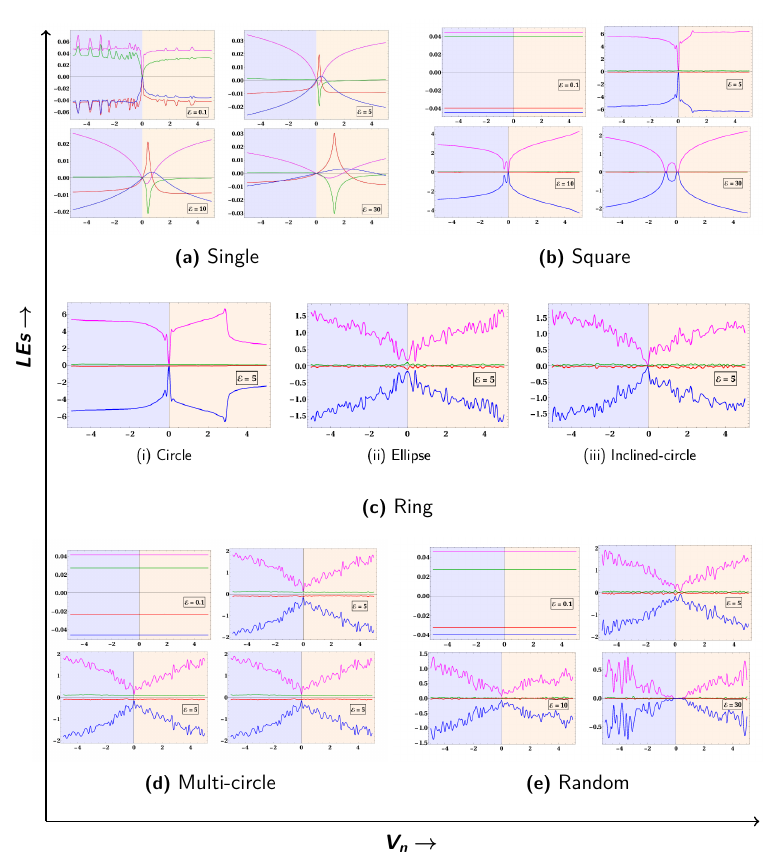}
	\caption{Lyapunov spectra of the Gaussian-perturbed Hamiltonian systems as functions of the control parameter $V_{n}$, representing the amplitude (perturbation strength) of the secondary Gaussian noise term, evaluated at different energies. The four Lyapunov exponents are shown in distinct colours (magenta, green, red, and blue). The region $V_{n}>0$ corresponds to the ``bump'' regime, while $V_{n}<0$ denotes the ``dimple'' regime. These two amplitude domains are visually distinguished by light pink (positive) and light blue (negative) background shading, respectively. }
	\label{LE-PLOT-VS-AMP}
\end{figure}
%%%%%%%%%%%%%%%%%%%%%%%%%%%%%%%%%%%

	To examine the energy dependence of instability, we evaluate the Lyapunov spectrum at four energy values: $\mathcal{E}=0.1,5,10,$ and $30$ by considering fixed amplitudes $V_{p}=5$ \& $V_{n}=2$. These values are chosen to probe the regimes ranging from strongly localized motion near the perturbation core to extended motion where the harmonic background dominates. The numerical results are summarized in \cref{table2}. 
	
	Across all geometrical configurations of perturbed landscapes, the largest Lyapunov exponent $\lambda_{1}$​ remains positive, confirming the presence of exponential sensitivity to initial conditions and hence deterministic chaos. However, the magnitude of $\lambda_{1}$ exhibits a clear and systematic dependence on the total energy.
	
	At very low energy ($\mathcal{E}=0.1$), although $\lambda_{1​}$ remains positive, its magnitude is comparatively small for all configurations. In this regime, trajectories are confined to a narrow region around the deformation centre, where motion is influenced by strong local curvature and there exist limited spatial exploration. The confinement restricts large-scale resonance interaction and reduces the effective phase-space transport rate. As a result, exponential divergence occurs, yet with relatively modest growth rates.
	
	At $\mathcal{E}=5$, $\lambda_{1}$​ maximizes across all perturbed configurations. This indicates that resonance overlap and nonlinear frequency modulation are most effective when trajectories explore those regions where the Gaussian deformation and the harmonic background contribute comparably to the effective curvature. In this regime, the interplay between localized perturbation and global confinement produces optimal conditions for instability and chaotic dynamics.
	
	With further increase in energy ($\mathcal{E}=10, 30$), the magnitude of $\lambda_{1}$​ decreases monotonically. At higher energies, trajectories explore larger spatial regions where the quadratic harmonic term dominates over the localized Gaussian contributions. Consequently, the system approaches  integrable limit of  unperturbed harmonic oscillator, and nonlinear resonance effects become comparatively weaker. Therefore, the diminishing $\lambda_{1}$  reflects a reduction in effective nonlinearity rather than the disappearance of chaos.
	
	Overall, the Lyapunov spectrum quantitatively supports the qualitative geometric evidence obtained from Poincaré maps: chaotic intensity is maximum at intermediate energies and decreases as total energy increases. The latter observation consistently  demonstrate the reduced relative impact of localized Gaussian deformations on the global phase-space structure. It is also noticed that for every perturbed configurations, the sum of all four exponents converges to zero after long times  for all considered energies. This result verifies the conservative (volume-preserving) character of the Gaussian-perturbed Hamiltonian systems. 
	
	%%%%%%%%%%%%%%%%%%%%%%%%%%%%%%%%%%%%%%%%%%%%%%%%%%%%%%%%%%%
	\subsubsection{\label{}Kaplan-Yorke Dimension ($D_{\mathrm{KY}}$):}
	%\subsection{\label{secIVd}Kaplan-Yorke Dimension: $D_{\mathrm{KY}}$}
	%%%%%%%%%%%%%%%%%%%%%%%%%%%%%%%%%%%%%%%%%%%%%%%%%%%%%%%%%%%%%%%%%
	This result, $D_{\mathrm{KY}}​=4$ for all chaotic dynamics observed in case of all perturbed model landscapes, is a generic feature of any $2D$ Hamiltonian systems and it is well understood from \cref{table2}.  Therefore, the $D_{\mathrm{KY}}$ does not provide additional discrimination in our case. The value $D_{\textrm{KY}}=4$ does not imply the presence of a strange attractor in phase space. Instead, it reflects a fundamental property of conservative systems i.e. (i) the chaotic motion occupies the entire accessible energy hypersurface, (ii) phase-space volume is preserved (Liouville’s theorem) and (iii) no dimensional collapse occurs. Therefore, for all Hamiltonian systems under consideration, the $D_{\mathrm{KY}}$ primarily serves as a consistency check rather than a discriminating measure of chaos. Hence, the largest Lyapunov exponent yet remains the most informative quantitative indicator of instability strength.
	
%%%%%%%%%%%%%%%%%%%%%%%%%%%%%%%%%%%%%%%%%%%%%%%%%
%	\begin{enumerate}[label={(\roman*)}]
%		\item The chaotic motion occupies the entire accessible energy hypersurface.
%		\item Phase-space volume is preserved (Liouville’s theorem).
%		\item No dimensional collapse occurs.
%	\end{enumerate}
%%%%%%%%%%%%%%%%%%%%%%%%%%%%%%%%%%%%%%%%%%%%%

	From  \cref{le-vs-E-plot},  we  notice that $\lambda_{1}$ is inversely proportional to energy for all perturbed models.  When we compare $\lambda_{1}$ values  at different energies (for fixed perturbation amplitudes), it is  observed that the degree of instability follows the approximate ordering  i.e.  $\lambda_{1,  \:\text{Square}}>\lambda_{1,  \: \text{Circle}}>\lambda_{1,  \:\text{Inclined-Circle}}>\lambda_{1,  \:\text{Multi-Circle}}>\lambda_{1, \: \text{Ellipse}}>\lambda_{1, \:\text{Random}}>\lambda_{1, \:\text{Single}}$.  Thus, the degree of chaos depends sensitively on geometric placement of Gaussian perturbations. In case of all configurations in which perturbations are positioned closer, these perturbations  together produce stronger effective deformation through Gaussian superposition, leading to enhanced resonance overlap and larger Lyapunov exponents. Therefore, spatial clustering plays a more decisive role than mere symmetry class.

	To quantify the role of noise,  Lyapunov spectra are computed by varying $V_n$ within the range of $-5$ to $5$, with $V_p$ fixed at $5$ for all Gaussian perturbed model systems and these spectra are illustrated in \cref{LE-PLOT-VS-AMP}. The constraint $V_n \leq V_p$ ensures that the central perturbation characterized by $V_p$ remains dominant. In the region $V_n>0$ (bump regime), the largest Lyapunov exponent increases approximately monotonically with $V_{n}$. A stronger positive Gaussian deformation increases local curvature and enhances nonlinear mode coupling, thereby amplifying exponential sensitivity to initial conditions. Similarly, in the region $V_n<0$ (dimple regime), increasing the magnitude $\abs{V_n}$ also increases $\lambda_{1}$. Although the sign of perturbed curvature differs in case of bumps and dimples, the absolute strength of local deformation governs resonance interaction. Hence, the instability depends primarily on perturbation magnitude rather than its sign. Thus, stronger localized deformations\textemdash whether repulsive (bumps) or attractive (dimples)\textemdash enhance dynamical instability. This symmetry in dynamical behaviour reflects the fact that instability is governed by curvature gradients rather than the sign of the potential modulation.
	
	From the Lyapunov spectra examined across all energies (see \cref{LE-PLOT-VS-AMP}), two consistent relationships emerge: $\lambda_{\max}\propto  \frac{\abs{V_{n}}}{\mathcal{E}}$. The first relation ($\propto V_{n}$) reflects direct amplification of nonlinear coupling when   perturbation strength of noise increases, while the second relation ($\propto \frac{1}{\mathcal{E}}$) expresses the diminishing relative influence of localized curvature on system dynamics when chaotic trajectories access higher-energy regions of phase space.

	%%%%%%%%%%%%%%%%%%%%%%%%%%%%%%%%%%%%%%%%%%%%%%%%%%%
	\section{\label{secVI}Conclusion}
	%%%%%%%%%%%%%%%%%
	In this study, we diagnose chaos in perturbed 2D harmonic potential wells, where the perturbation in the system's Hamiltonian is introduced by two Gaussian-like terms. The primary perturbation arises due to a central single Gaussian bump/dimple inside the well, and the secondary perturbation arises due to the distribution of multiple Gaussian bumps/dimples (noise) in various geometrical shapes. To carry out this analysis, we need to construct various perturbed model landscapes, in which the geometrical arrangements of noise (multiple secondary perturbations) are inspired by several cymatics patterns. We consider square, circular, elliptical, inclined-circular, multi-circular and random arrangements of Gaussian noise along with a single Gaussian perturbation placed at the centre of the harmonic potential well to construct the perturbed models. In the case of all considered models, the amplitude of the primary perturbation term (central single bump) is always greater than that of the secondary perturbation terms (noise). In our study, we keep the width of the Gaussian perturbation term fixed. We implement various chaos diagnostic tools, namely, time-series plots, phase-space trajectories, power spectra, auto-correlation functions, Poincaré maps, Lyapunov exponents to characterize the chaotic dynamics in all the Gaussian perturbed model systems.

	We conclude that the chaotic dynamics emerging from our study are intricately linked to the properties of the perturbations applied to our model systems. Specifically, these dynamics are directly proportional to the bump amplitude and inversely proportional to the bump width in a single-bump perturbed model. Furthermore, across various perturbed landscapes, we have determined that the chaotic behavior observed in different model systems at varying energy levels is similarly dependent: direct proportionality to the amplitude of Gaussian noise and inverse dependence on energy values. These compelling results are firmly validated through rigorous Poincaré map analyses and calculations of Lyapunov exponents.
	
	The Lyapunov exponent $\lambda$ provides a quantitative measure of the system's sensitivity to initial conditions, allowing us to map the transition from regular motion ($\lambda=0$) to chaotic motion ($\lambda>0$) as a function of amplitude and width of Gaussian perturbation , as well as the energy $\mathcal{E}$. Complementing this quantitative measure, we construct Poincaré map to visualize the topology of the phase space. These Poincaré maps allow us to understand the intricate coexistence of regular islands, invariant tori, and the chaotic sea in phase space, which characterizes the mixed dynamics of our model systems. 
	
	Importantly, our findings demonstrate that the cymatic-inspired noise distribution acts as a significant control parameter, fundamentally influencing the pathway to chaos. We can confidently state that the chaos observed in Gaussian perturbed harmonic potential wells is predominantly driven by noise. This research not only offers a physically motivated and mathematically manageable family of potentials but also serves as a comprehensive guide for probing the complex dynamics in perturbed systems where order and disorder are inextricably intertwined.
	
%%%%%%%%%%%%%%%%%%%%%%%%%%%%%%%%%%%%%
%%%%%%%%%%%%%%%%%%%%%%%%%%%%%%%%%%%%%%%%%%%%%%%%%%%%%%
%\begin{acknowledgments}
%\end{acknowledgments}
%%%%%%%%%%%%%%%%%%%%%%%%%%%%%%%%
\section*{Data availability statement}
No data available. The data that support the findings of this study are available upon reasonable request from
the authors.

%%%%%%%%%%%%%%%%%%%%%%%%%%%%%%%%
\section*{Declaration of competing interests}
The authors declare that they have no competing financial interests or personal relationships that could
potentially affect the integrity of the work presented in this paper.

%%%%%%%%%%%%%%%%%%%%%%%%%%%%%%%%
\section*{Author contributions}
T.P. and P.P.D. conceptualized the work, developed the methodology; T.P. implemented the software, curated the data, and prepared visualizations; P.P.D. and T.P. performed the formal analysis and drafted the manuscript; T.P., P.P.D. and B.G. validated the results, reviewed, and edited the manuscript; B.G. supervised and administered the work.
%%%%%%%%%%%%%%%%%%%%%%%%%%%%%%%%%%%%%%%%%%%%%%%%
\appendix
%\section{\label{sec7}Appendix}

%%%%%%%%%%%%%%%%%%%%%%%%%%%%%%%%%%%%%%%%%%%%%%%%%

\bibliography{References}

\begin{thebibliography}{22}%
\makeatletter
\providecommand \@ifxundefined [1]{%
 \@ifx{#1\undefined}
}%
\providecommand \@ifnum [1]{%
 \ifnum #1\expandafter \@firstoftwo
 \else \expandafter \@secondoftwo
 \fi
}%
\providecommand \@ifx [1]{%
 \ifx #1\expandafter \@firstoftwo
 \else \expandafter \@secondoftwo
 \fi
}%
\providecommand \natexlab [1]{#1}%
\providecommand \enquote  [1]{``#1''}%
\providecommand \bibnamefont  [1]{#1}%
\providecommand \bibfnamefont [1]{#1}%
\providecommand \citenamefont [1]{#1}%
\providecommand \href@noop [0]{\@secondoftwo}%
\providecommand \href [0]{\begingroup \@sanitize@url \@href}%
\providecommand \@href[1]{\@@startlink{#1}\@@href}%
\providecommand \@@href[1]{\endgroup#1\@@endlink}%
\providecommand \@sanitize@url [0]{\catcode `\\12\catcode `\$12\catcode
  `\&12\catcode `\#12\catcode `\^12\catcode `\_12\catcode `\%12\relax}%
\providecommand \@@startlink[1]{}%
\providecommand \@@endlink[0]{}%
\providecommand \url  [0]{\begingroup\@sanitize@url \@url }%
\providecommand \@url [1]{\endgroup\@href {#1}{\urlprefix }}%
\providecommand \urlprefix  [0]{URL }%
\providecommand \Eprint [0]{\href }%
\providecommand \doibase [0]{https://doi.org/}%
\providecommand \selectlanguage [0]{\@gobble}%
\providecommand \bibinfo  [0]{\@secondoftwo}%
\providecommand \bibfield  [0]{\@secondoftwo}%
\providecommand \translation [1]{[#1]}%
\providecommand \BibitemOpen [0]{}%
\providecommand \bibitemStop [0]{}%
\providecommand \bibitemNoStop [0]{.\EOS\space}%
\providecommand \EOS [0]{\spacefactor3000\relax}%
\providecommand \BibitemShut  [1]{\csname bibitem#1\endcsname}%
\let\auto@bib@innerbib\@empty
%</preamble>
\bibitem [{\citenamefont {Poincar{\'e}}(2017)}]{poincare2017three}%
  \BibitemOpen
  \bibfield  {author} {\bibinfo {author} {\bibfnamefont {H.}~\bibnamefont
  {Poincar{\'e}}},\ }\href@noop {} {\emph {\bibinfo {title} {The three-body
  problem and the equations of dynamics: Poincar{\'e}’s foundational work on
  dynamical systems theory}}}\ (\bibinfo  {publisher} {Springer},\ \bibinfo
  {year} {2017})\BibitemShut {NoStop}%
\bibitem [{\citenamefont {Arnol'd}(1963)}]{Vladimir_I_Arnol'd_1963}%
  \BibitemOpen
  \bibfield  {author} {\bibinfo {author} {\bibfnamefont {V.~I.}\ \bibnamefont
  {Arnol'd}},\ }\bibfield  {title} {\bibinfo {title} {Small denominators and
  problems of stability of motion in classical and celestial mechanics},\
  }\href {https://doi.org/10.1070/RM1963v018n06ABEH001143} {\bibfield
  {journal} {\bibinfo  {journal} {Russian Mathematical Surveys}\ }\textbf
  {\bibinfo {volume} {18}},\ \bibinfo {pages} {85} (\bibinfo {year}
  {1963})}\BibitemShut {NoStop}%
\bibitem [{\citenamefont {Fishman}\ \emph {et~al.}(1982)\citenamefont
  {Fishman}, \citenamefont {Grempel},\ and\ \citenamefont
  {Prange}}]{PhysRevLett.49.509}%
  \BibitemOpen
  \bibfield  {author} {\bibinfo {author} {\bibfnamefont {S.}~\bibnamefont
  {Fishman}}, \bibinfo {author} {\bibfnamefont {D.~R.}\ \bibnamefont
  {Grempel}},\ and\ \bibinfo {author} {\bibfnamefont {R.~E.}\ \bibnamefont
  {Prange}},\ }\bibfield  {title} {\bibinfo {title} {Chaos, quantum
  recurrences, and anderson localization},\ }\href
  {https://doi.org/10.1103/PhysRevLett.49.509} {\bibfield  {journal} {\bibinfo
  {journal} {Phys. Rev. Lett.}\ }\textbf {\bibinfo {volume} {49}},\ \bibinfo
  {pages} {509} (\bibinfo {year} {1982})}\BibitemShut {NoStop}%
\bibitem [{\citenamefont {Geisel}\ \emph {et~al.}(1991)\citenamefont {Geisel},
  \citenamefont {Ketzmerick},\ and\ \citenamefont
  {Petschel}}]{PhysRevLett.67.3635}%
  \BibitemOpen
  \bibfield  {author} {\bibinfo {author} {\bibfnamefont {T.}~\bibnamefont
  {Geisel}}, \bibinfo {author} {\bibfnamefont {R.}~\bibnamefont {Ketzmerick}},\
  and\ \bibinfo {author} {\bibfnamefont {G.}~\bibnamefont {Petschel}},\
  }\bibfield  {title} {\bibinfo {title} {Metamorphosis of a cantor spectrum due
  to classical chaos},\ }\href {https://doi.org/10.1103/PhysRevLett.67.3635}
  {\bibfield  {journal} {\bibinfo  {journal} {Phys. Rev. Lett.}\ }\textbf
  {\bibinfo {volume} {67}},\ \bibinfo {pages} {3635} (\bibinfo {year}
  {1991})}\BibitemShut {NoStop}%
\bibitem [{\citenamefont {Chladni}(1787)}]{chladni1787entdeckungen}%
  \BibitemOpen
  \bibfield  {author} {\bibinfo {author} {\bibfnamefont {E.~F.~F.}\
  \bibnamefont {Chladni}},\ }\href@noop {} {\emph {\bibinfo {title}
  {Entdeckungen {\"u}ber die Theorie des Klanges}}}\ (\bibinfo  {publisher}
  {Zentralantiquariat der Deutschen Demokratischen Republik},\ \bibinfo {year}
  {1787})\BibitemShut {NoStop}%
\bibitem [{\citenamefont {Chladni}(1967)}]{BB10515802}%
  \BibitemOpen
  \bibfield  {author} {\bibinfo {author} {\bibfnamefont {E.~F.~F.}\
  \bibnamefont {Chladni}},\ }\href {https://ci.nii.ac.jp/ncid/BB10515802}
  {\emph {\bibinfo {title} {Entdeckungen uber die Theorie des Klanges}}},\
  Landmarks of science / ed. by Harold Hartley, Duane H. D. Roller\ (\bibinfo
  {publisher} {Readex Microprint},\ \bibinfo {year} {1967})\BibitemShut
  {NoStop}%
\bibitem [{\citenamefont {Bell}(1969)}]{bell1969gaussian}%
  \BibitemOpen
  \bibfield  {author} {\bibinfo {author} {\bibfnamefont {S.}~\bibnamefont
  {Bell}},\ }\bibfield  {title} {\bibinfo {title} {Gaussian perturbations to
  harmonic oscillators},\ }\href@noop {} {\bibfield  {journal} {\bibinfo
  {journal} {Journal of Physics B: Atomic and Molecular Physics}\ }\textbf
  {\bibinfo {volume} {2}},\ \bibinfo {pages} {1001} (\bibinfo {year}
  {1969})}\BibitemShut {NoStop}%
\bibitem [{\citenamefont {Earl}(2008)}]{earl2008harmonic}%
  \BibitemOpen
  \bibfield  {author} {\bibinfo {author} {\bibfnamefont {B.~L.}\ \bibnamefont
  {Earl}},\ }\bibfield  {title} {\bibinfo {title} {The harmonic oscillator with
  a gaussian perturbation: evaluation of the integrals and example
  applications},\ }\href@noop {} {\bibfield  {journal} {\bibinfo  {journal}
  {Journal of chemical education}\ }\textbf {\bibinfo {volume} {85}},\ \bibinfo
  {pages} {453} (\bibinfo {year} {2008})}\BibitemShut {NoStop}%
\bibitem [{\citenamefont {Luukko}\ \emph {et~al.}(2016)\citenamefont {Luukko},
  \citenamefont {Drury}, \citenamefont {Klales}, \citenamefont {Kaplan},
  \citenamefont {Heller},\ and\ \citenamefont {R{\"a}s{\"a}nen}}]{Luukko2016}%
  \BibitemOpen
  \bibfield  {author} {\bibinfo {author} {\bibfnamefont {P.~J.~J.}\
  \bibnamefont {Luukko}}, \bibinfo {author} {\bibfnamefont {B.}~\bibnamefont
  {Drury}}, \bibinfo {author} {\bibfnamefont {A.}~\bibnamefont {Klales}},
  \bibinfo {author} {\bibfnamefont {L.}~\bibnamefont {Kaplan}}, \bibinfo
  {author} {\bibfnamefont {E.~J.}\ \bibnamefont {Heller}},\ and\ \bibinfo
  {author} {\bibfnamefont {E.}~\bibnamefont {R{\"a}s{\"a}nen}},\ }\bibfield
  {title} {\bibinfo {title} {Strong quantum scarring by local impurities},\
  }\href {https://doi.org/10.1038/srep37656} {\bibfield  {journal} {\bibinfo
  {journal} {Scientific Reports}\ }\textbf {\bibinfo {volume} {6}},\ \bibinfo
  {pages} {37656} (\bibinfo {year} {2016})}\BibitemShut {NoStop}%
\bibitem [{\citenamefont {Keski-Rahkonen}\ \emph {et~al.}(2019)\citenamefont
  {Keski-Rahkonen}, \citenamefont {Luukko}, \citenamefont {Åberg},\ and\
  \citenamefont {Räsänen}}]{Keski-Rahkonen_2019}%
  \BibitemOpen
  \bibfield  {author} {\bibinfo {author} {\bibfnamefont {J.}~\bibnamefont
  {Keski-Rahkonen}}, \bibinfo {author} {\bibfnamefont {P.~J.~J.}\ \bibnamefont
  {Luukko}}, \bibinfo {author} {\bibfnamefont {S.}~\bibnamefont {Åberg}},\
  and\ \bibinfo {author} {\bibfnamefont {E.}~\bibnamefont {Räsänen}},\
  }\bibfield  {title} {\bibinfo {title} {Effects of scarring on quantum chaos
  in disordered quantum wells},\ }\href
  {https://doi.org/10.1088/1361-648X/aaf9fb} {\bibfield  {journal} {\bibinfo
  {journal} {Journal of Physics: Condensed Matter}\ }\textbf {\bibinfo {volume}
  {31}},\ \bibinfo {pages} {105301} (\bibinfo {year} {2019})}\BibitemShut
  {NoStop}%
\bibitem [{\citenamefont {Selinummi}\ \emph {et~al.}(2024)\citenamefont
  {Selinummi}, \citenamefont {Keski-Rahkonen}, \citenamefont {Chalangari},\
  and\ \citenamefont {R\"as\"anen}}]{PhysRevB.110.235420}%
  \BibitemOpen
  \bibfield  {author} {\bibinfo {author} {\bibfnamefont {S.}~\bibnamefont
  {Selinummi}}, \bibinfo {author} {\bibfnamefont {J.}~\bibnamefont
  {Keski-Rahkonen}}, \bibinfo {author} {\bibfnamefont {F.}~\bibnamefont
  {Chalangari}},\ and\ \bibinfo {author} {\bibfnamefont {E.}~\bibnamefont
  {R\"as\"anen}},\ }\bibfield  {title} {\bibinfo {title} {Formation,
  prevalence, and stability of bouncing-ball quantum scars},\ }\href
  {https://doi.org/10.1103/PhysRevB.110.235420} {\bibfield  {journal} {\bibinfo
   {journal} {Phys. Rev. B}\ }\textbf {\bibinfo {volume} {110}},\ \bibinfo
  {pages} {235420} (\bibinfo {year} {2024})}\BibitemShut {NoStop}%
\bibitem [{\citenamefont {Amore}\ \emph {et~al.}(2024)\citenamefont {Amore},
  \citenamefont {Fern{\'a}ndez},\ and\ \citenamefont {Garcia}}]{Amore2024}%
  \BibitemOpen
  \bibfield  {author} {\bibinfo {author} {\bibfnamefont {P.}~\bibnamefont
  {Amore}}, \bibinfo {author} {\bibfnamefont {F.~M.}\ \bibnamefont
  {Fern{\'a}ndez}},\ and\ \bibinfo {author} {\bibfnamefont {J.}~\bibnamefont
  {Garcia}},\ }\bibfield  {title} {\bibinfo {title} {On the eigenvalues of the
  harmonic oscillator with a gaussian perturbation},\ }\href
  {https://doi.org/10.1140/epjp/s13360-024-05736-5} {\bibfield  {journal}
  {\bibinfo  {journal} {The European Physical Journal Plus}\ }\textbf {\bibinfo
  {volume} {139}},\ \bibinfo {pages} {920} (\bibinfo {year}
  {2024})}\BibitemShut {NoStop}%
\bibitem [{\citenamefont {Fassari}\ \emph {et~al.}(2020)\citenamefont
  {Fassari}, \citenamefont {Nieto},\ and\ \citenamefont
  {Rinaldi}}]{Fassari2020}%
  \BibitemOpen
  \bibfield  {author} {\bibinfo {author} {\bibfnamefont {S.}~\bibnamefont
  {Fassari}}, \bibinfo {author} {\bibfnamefont {L.~M.}\ \bibnamefont {Nieto}},\
  and\ \bibinfo {author} {\bibfnamefont {F.}~\bibnamefont {Rinaldi}},\
  }\bibfield  {title} {\bibinfo {title} {The two lowest eigenvalues of the
  harmonic oscillator in the presence of a gaussian perturbation},\ }\href
  {https://doi.org/10.1140/epjp/s13360-020-00761-6} {\bibfield  {journal}
  {\bibinfo  {journal} {The European Physical Journal Plus}\ }\textbf {\bibinfo
  {volume} {135}},\ \bibinfo {pages} {728} (\bibinfo {year}
  {2020})}\BibitemShut {NoStop}%
\bibitem [{\citenamefont {Holmes}(1990)}]{HOLMES1990137}%
  \BibitemOpen
  \bibfield  {author} {\bibinfo {author} {\bibfnamefont {P.}~\bibnamefont
  {Holmes}},\ }\bibfield  {title} {\bibinfo {title} {Poincaré, celestial
  mechanics, dynamical-systems theory and “chaos”},\ }\href
  {https://doi.org/https://doi.org/10.1016/0370-1573(90)90012-Q} {\bibfield
  {journal} {\bibinfo  {journal} {Physics Reports}\ }\textbf {\bibinfo {volume}
  {193}},\ \bibinfo {pages} {137} (\bibinfo {year} {1990})}\BibitemShut
  {NoStop}%
\bibitem [{\citenamefont {Shahhosseini}\ \emph {et~al.}(2023)\citenamefont
  {Shahhosseini}, \citenamefont {Tien},\ and\ \citenamefont
  {D’Souza}}]{shahhosseini2023poincare}%
  \BibitemOpen
  \bibfield  {author} {\bibinfo {author} {\bibfnamefont {A.}~\bibnamefont
  {Shahhosseini}}, \bibinfo {author} {\bibfnamefont {M.-H.}\ \bibnamefont
  {Tien}},\ and\ \bibinfo {author} {\bibfnamefont {K.}~\bibnamefont
  {D’Souza}},\ }\bibfield  {title} {\bibinfo {title} {Poincare maps: a modern
  systematic approach toward obtaining effective sections},\ }\href@noop {}
  {\bibfield  {journal} {\bibinfo  {journal} {Nonlinear Dynamics}\ }\textbf
  {\bibinfo {volume} {111}},\ \bibinfo {pages} {529} (\bibinfo {year}
  {2023})}\BibitemShut {NoStop}%
\bibitem [{\citenamefont {ABARBANEL}\ \emph {et~al.}(1991)\citenamefont
  {ABARBANEL}, \citenamefont {BROWN},\ and\ \citenamefont
  {KENNEL}}]{doi:10.1142/S021797929100064X}%
  \BibitemOpen
  \bibfield  {author} {\bibinfo {author} {\bibfnamefont {H.~D.~I.}\
  \bibnamefont {ABARBANEL}}, \bibinfo {author} {\bibfnamefont {R.}~\bibnamefont
  {BROWN}},\ and\ \bibinfo {author} {\bibfnamefont {M.~B.}\ \bibnamefont
  {KENNEL}},\ }\bibfield  {title} {\bibinfo {title} {Lyapunov exponents in
  chaotic systems: Their importance and their evaluation using observed data},\
  }\href {https://doi.org/10.1142/S021797929100064X} {\bibfield  {journal}
  {\bibinfo  {journal} {International Journal of Modern Physics B}\ }\textbf
  {\bibinfo {volume} {05}},\ \bibinfo {pages} {1347} (\bibinfo {year}
  {1991})},\ \Eprint
  {https://arxiv.org/abs/https://doi.org/10.1142/S021797929100064X}
  {https://doi.org/10.1142/S021797929100064X} \BibitemShut {NoStop}%
\bibitem [{\citenamefont {Skokos}(2009)}]{2009}%
  \BibitemOpen
  \bibfield  {author} {\bibinfo {author} {\bibfnamefont {C.}~\bibnamefont
  {Skokos}},\ }\bibfield  {title} {\bibinfo {title} {The lyapunov
  characteristic exponents and their computation},\ }\href
  {https://doi.org/10.1007/978-3-642-04458-8_2} {\bibfield  {journal} {\bibinfo
   {journal} {Lecture Notes in Physics}\ ,\ \bibinfo {pages} {63–135}}
  (\bibinfo {year} {2009})}\BibitemShut {NoStop}%
\bibitem [{\citenamefont {Christiansen}\ and\ \citenamefont
  {Rugh}(1997)}]{christiansen1997computing}%
  \BibitemOpen
  \bibfield  {author} {\bibinfo {author} {\bibfnamefont {F.}~\bibnamefont
  {Christiansen}}\ and\ \bibinfo {author} {\bibfnamefont {H.~H.}\ \bibnamefont
  {Rugh}},\ }\bibfield  {title} {\bibinfo {title} {Computing lyapunov spectra
  with continuous gram-schmidt orthonormalization},\ }\href@noop {} {\bibfield
  {journal} {\bibinfo  {journal} {Nonlinearity}\ }\textbf {\bibinfo {volume}
  {10}},\ \bibinfo {pages} {1063} (\bibinfo {year} {1997})}\BibitemShut
  {NoStop}%
\bibitem [{\citenamefont {Evans}\ \emph {et~al.}(2000)\citenamefont {Evans},
  \citenamefont {Cohen}, \citenamefont {Searles},\ and\ \citenamefont
  {Bonetto}}]{Evans2000}%
  \BibitemOpen
  \bibfield  {author} {\bibinfo {author} {\bibfnamefont {D.~J.}\ \bibnamefont
  {Evans}}, \bibinfo {author} {\bibfnamefont {E.~G.~D.}\ \bibnamefont {Cohen}},
  \bibinfo {author} {\bibfnamefont {D.~J.}\ \bibnamefont {Searles}},\ and\
  \bibinfo {author} {\bibfnamefont {F.}~\bibnamefont {Bonetto}},\ }\bibfield
  {title} {\bibinfo {title} {Note on the kaplan--yorke dimension and linear
  transport coefficients},\ }\href {https://doi.org/10.1023/A:1026449702528}
  {\bibfield  {journal} {\bibinfo  {journal} {Journal of Statistical Physics}\
  }\textbf {\bibinfo {volume} {101}},\ \bibinfo {pages} {17} (\bibinfo {year}
  {2000})}\BibitemShut {NoStop}%
\bibitem [{\citenamefont {Chen}\ \emph {et~al.}(2018)\citenamefont {Chen},
  \citenamefont {Bayani}, \citenamefont {Akgul}, \citenamefont {Jafari},
  \citenamefont {Pham}, \citenamefont {Wang},\ and\ \citenamefont
  {Jafari}}]{Chen2018}%
  \BibitemOpen
  \bibfield  {author} {\bibinfo {author} {\bibfnamefont {H.}~\bibnamefont
  {Chen}}, \bibinfo {author} {\bibfnamefont {A.}~\bibnamefont {Bayani}},
  \bibinfo {author} {\bibfnamefont {A.}~\bibnamefont {Akgul}}, \bibinfo
  {author} {\bibfnamefont {M.-A.}\ \bibnamefont {Jafari}}, \bibinfo {author}
  {\bibfnamefont {V.-T.}\ \bibnamefont {Pham}}, \bibinfo {author}
  {\bibfnamefont {X.}~\bibnamefont {Wang}},\ and\ \bibinfo {author}
  {\bibfnamefont {S.}~\bibnamefont {Jafari}},\ }\bibfield  {title} {\bibinfo
  {title} {A flexible chaotic system with adjustable amplitude, largest
  lyapunov exponent, and local kaplan--yorke dimension and its usage in
  engineering applications},\ }\href
  {https://doi.org/10.1007/s11071-018-4162-9} {\bibfield  {journal} {\bibinfo
  {journal} {Nonlinear Dynamics}\ }\textbf {\bibinfo {volume} {92}},\ \bibinfo
  {pages} {1791} (\bibinfo {year} {2018})}\BibitemShut {NoStop}%
\bibitem [{\citenamefont {Ma}\ and\ \citenamefont
  {Huang}(2022)}]{doi:10.1142/S0218127422502224}%
  \BibitemOpen
  \bibfield  {author} {\bibinfo {author} {\bibfnamefont {L.}~\bibnamefont
  {Ma}}\ and\ \bibinfo {author} {\bibfnamefont {C.}~\bibnamefont {Huang}},\
  }\bibfield  {title} {\bibinfo {title} {Comparative analysis of correlation
  and kaplan–yorke dimensions for discrete-time fractional systems},\ }\href
  {https://doi.org/10.1142/S0218127422502224} {\bibfield  {journal} {\bibinfo
  {journal} {International Journal of Bifurcation and Chaos}\ }\textbf
  {\bibinfo {volume} {32}},\ \bibinfo {pages} {2250222} (\bibinfo {year}
  {2022})},\ \Eprint
  {https://arxiv.org/abs/https://doi.org/10.1142/S0218127422502224}
  {https://doi.org/10.1142/S0218127422502224} \BibitemShut {NoStop}%
\bibitem [{\citenamefont {Chlouverakis}\ and\ \citenamefont
  {Sprott}(2005)}]{CHLOUVERAKIS2005156}%
  \BibitemOpen
  \bibfield  {author} {\bibinfo {author} {\bibfnamefont {K.~E.}\ \bibnamefont
  {Chlouverakis}}\ and\ \bibinfo {author} {\bibfnamefont {J.}~\bibnamefont
  {Sprott}},\ }\bibfield  {title} {\bibinfo {title} {A comparison of
  correlation and lyapunov dimensions},\ }\href
  {https://doi.org/https://doi.org/10.1016/j.physd.2004.10.006} {\bibfield
  {journal} {\bibinfo  {journal} {Physica D: Nonlinear Phenomena}\ }\textbf
  {\bibinfo {volume} {200}},\ \bibinfo {pages} {156} (\bibinfo {year}
  {2005})}\BibitemShut {NoStop}%
\end{thebibliography}%
%\nocite{*}

\end{document}